\documentclass[aps,twocolumn,nofootinbib,showpacs,final,balancelastpage,superscriptaddress]{revtex4}
\usepackage{graphicx}

\usepackage{tabularx}
\usepackage{amssymb}
\usepackage{amsmath}
\usepackage{amsbsy}
\usepackage{comment}
\usepackage{mathrsfs}
\usepackage{dsfont}
\usepackage{nicefrac}
\usepackage{wrapfig}
\usepackage{amsfonts}
\usepackage{url} 
\usepackage[nolist]{acronym}
\usepackage{nicefrac}
\usepackage{subfigure}
\usepackage{epsf,color,colordvi,pifont}
\usepackage{showkeys}
\DeclareFontFamily{OT1}{pzc}{}
\DeclareFontShape{OT1}{pzc}{m}{it}{<-> s * [1.10] pzcmi7t}{}
\DeclareMathAlphabet{\mathpzc}{OT1}{pzc}{m}{it}
\def\dbar{{\mathchar'26\mkern-12mu \textrm{d}}}
\def\deltabar{{\mathchar'26\mkern-9mu\delta}}

\begin{document}
\title{Signatures of the Schwinger mechanism assisted by a fast-oscillating electric field}
\author{Selym \surname{Villalba-Ch\'avez}}
\email{selym@tp1.uni-duesseldorf.de}
\author{Carsten  \surname{M\"{u}ller}}
\affiliation{Institut f\"{u}r Theoretische Physik, Heinrich-Heine-Universit\"{a}t D\"{u}sseldorf,\\ Universit\"{a}tsstr.\,1, 40225 D\"{u}sseldorf, Germany}

\begin{abstract}
The spontaneous production of electron-positron pairs from the vacuum--in  a field configuration composed of a high-frequency electric mode of weak intensity and 
a strong constant electric field--is investigated. Asymptotic expressions for the single-particle distribution function ruling this nonperturbative process are 
established by considering the low-density approximation in the Boltzmann-Vlasov equation.  An analytical formula for the density rate of yielded particles is 
established which is shown to  manifest a nonperturbative dependence on both the strong and weak electric fields and to interpolate between the tunneling and 
multi-photon regimes. It is shown that--under appropriate circumstances--the produced plasma of electrons and positrons might reach densities for which their 
recombinations into high-energy photons occurs copiously. On the basis of this  feature, an experimental setup for observing the dynamically-assisted Schwinger effect is put forward.
\end{abstract}



\date{\today}

\maketitle

\section{Introduction}

Finding suitable  and controllable experimental conditions to materialize the all-permeating quantum vacuum fluctuations has been a fundamental goal in particle 
physics since the time when our inert vacuum perception strikly changed into a nontrivial regulatory void, responsible of mediating the  interactions between elementary 
particles. Even before the full establishment  of quantum electrodynamics (QED), it was noted that this sort of vacuum instability could 
be conceived by producing electron-positron pairs if  a macroscopic electric field  $E$ is held in vacuum \cite{Sauter:1931,Heisenberg:1935,Schwinger:1951nm}.  
Notwithstanding, the corresponding pair production (PP) rate $\mathcal{R}\sim \exp[-\pi E_{\mathrm{cr}}/E]$ provides evidences that an experimental 
verification of this so-called Schwinger mechanism is far from our reach; save yet unaccessible field strengths--comparable to the critical 
scale of QED $E_{\mathrm{cr}}=m^2/e\sim 10^{16}\ \rm V/cm$--become available.\footnote{Here and henceforth the mass and the absolute 
charge of an electron will be denoted by $m$ and $e$, respectively. Besides, throughout the manuscript Heaviside$-$Lorentz units--with 
the speed of light and the Planck constant set to unity $c=\hbar =1$--are used.} Although significant progresses toward high-intensity laser 
technology are raising our hopes of reaching the required field strengths within the focal spot of multipetawatt laser pulses, it is rather 
likely that an experimental verification of the Schwinger PP process remains a challenging task to achieve, at least in a near future. Mainly, 
because the  peak field strengths $\sim 10^{-2}E_{\mathrm{cr}}$ expected at the new generation of laser systems, including the Extreme Light 
Infrastructure (ELI) \cite{ELI} and  the Exawatt Center for Extreme Light Studies (XCELS) \cite{xcels}, 
would keep the production rate  very small. 

A central aspect in investigations  aiming to relieve the exponential suppression of $\mathcal{R}$ is the identification of field setups which 
may allow us to maximize the Schwinger effect \cite{Hebenstreit:2009km,Bulanov,BlaschkeCPP,Kohlfurst:2012rb,Hebenstreit:2014lra,Gonoskov}. Perhaps 
the most robust configuration found so far is the one implemented in what is nowadays known as the dynamically-assisted Schwinger mechanism \cite{dgs2008,dgs2009}, 
where--in addition to a strong quasi-static electric field--a weak but high-frequency field component is superimposed. In the original papers on the subject, 
the combined field was composed  of two Sauter pulses \cite{dgs2008} or a constant electric field and a high energy electromagnetic wave  with  
$\omega\lesssim 2m$ \cite{dgs2009}. The latter ingredient--partially motivated by the experimental verification of the nonlinear Breit-Wheeler 
reaction \cite{Burke:1997ew}--stimulates the creation of pairs substantially. Indeed, first estimates resulting from this assisted scenario predict 
an enhancement of the PP rate $\mathcal{R}\sim \exp[-\kappa E_{\rm cr}/E]$ with $0<\kappa\ll1$,  while its nonperturbative feature in the strong field 
strength is kept. Similar improvement has been predicted to take place in production channels other than the one described so far, provided the assisted 
high-frequency laser wave is  present \cite{DiPiazza:2009py,Jansen2013,Augustin}. Qualitatively, this sort of catalysis is understood as a direct 
consequence of the absorption of photons from the weak field, which causes an effective reduction of the barrier width that an electron has to tunnel 
from  negative to positive Dirac continuum. A large number of transitions are thus facilitated--pairs are created copiously--leading to increase 
our chances for  observing a  signature of the vacuum instability.

This paper is devoted to study the spontaneous production of electron-positron pairs as might occur in a dynamically-assisted setup driven by combination of 
a constant and a purely  time-dependent electric field. Our theoretical approach relies on the quantum transport equation that dictates the time evolution of 
the PP process \cite{Schmidt:1998vi,Kluger,Schmidt:1999vi,alkofer:2001ib}. Noteworthy, several investigations of this nature have already been carried 
out; most of them by using numerical techniques from which valuable information and features have been extracted \cite{Orthaber2011,Grobe2012,Akal:2014eua,Aleksandrov:2018uqb,Panferov:2015yda,Otto:2014ssa}.  
Meanwhile various research have focused on deriving formulae for the created particles spectra \cite{Linder:2015vta,Otto:2014ssa,Panferov:2015yda,torgrimsson2016,Fey}. 
This way, illuminating the crucial aspects from which an optimized version of the aforementioned enhancement could be reached. Two recent 
papers went a step further by providing  analytical expressions for the total probability of produced pairs \cite{Togrimsson1,Togrimsson2}. In these investigations 
a pertubative treatment in the weak field was used within the WKB and the world-line instanton methods, respectively. Particular attention  was laid on  
weak fields with Sauter and Gaussian profiles. However, the case of a periodically oscillating  mode was touched only  briefly. Here we complement these ana\-lytical studies  
by using a quantum kinetic approach in which both the weak and strong field are treated  nonperturbatively. We discuss in details the case in which the assisted mode 
oscillates periodically and obtain a formula for the density rate of yielded particles which is shown to  manifest a nonperturbative dependence on both the strong 
and weak electric fields and to interpolate between the tunneling and multi-photon regimes. Besides, the outcomes of this analysis are exploited to  put forward an 
experimental setup which aims to verify indirectly the realization of the  dynamically-assisted Schwinger effect. We show that--once the field is switched off--the 
plasma density can be  high enough as to facilitate the copious annihilation of  pairs into gamma photons. The detection of these photons is the central aspect 
in the proposed setup. 

This paper is organized as follows: In Sec.~\ref{generalasp} we adopt the model to be analyzed and summarize briefly the main aspects linked to 
the  quantum Vlasov equation and its solution within the low-density limit. The aforementioned approximation is particularized in Sec.~\ref{SPDFasymptotic} 
to the case in which the external background combines a strong static electric field and a fast-oscillating electric field. There we establish an 
analytic expression for the single-particle distribution function and discuss its behavior in various regimes of interest. Later on, in Sec.~\ref{observabledcp}, 
we derive a compact asymptotic formula for the PP rate and reveal explicitly how the enhancement caused by the weak mode is closely connected with 
the perturbative PP rate associated with the absorption of photons. Details about some  special aspects of these calculations are given in Appendices~\ref{AppendixA} 
and \ref{AppendixB}. Finally, in Sec.~\ref{emissionsec}, we investigate  the evolution of the electron-positron plasma after its creation via  
the assisted Schwinger mechanism. An expression for the total number of photons resulting from the annihilation  of pairs at early times is  derived.  
Some  insight on a  viable experimental setup aiming to detect these photons,  and thus the spontaneous production of pairs, is given afterwards.

\section{General aspects \label{generalasp}}

We consider the spontaneous production of electron-positron pairs taking place in a time-dependent but homogeneous electric field combining a strong and a weak mode 
with frequencies $\Omega$ and $\omega$, respectively. We will suppose this field is localized temporally between $-T/2\leqslant t\leqslant T/2$, its pulse length $T=2\pi N /\omega$ 
being determined by the number of cycles $N$ and $\omega$. In the following,  we further restrict the model to the case in which the variation linked to the 
perturbative mode is much faster than the one undergone by the strong field counterpart [$\omega\gg\Omega$] and where, accordingly, $N$ is very large. Hence, the PP problem can be formulated as if the 
creation process was taking place in a background characterized by a constant electric field and a fast-oscillating mode generated by the four-potential
\begin{equation}
\begin{split}
\mathpzc{A}^\mu(t)&=-\left[E_s t+\frac{E_w}{\omega}\sin(\omega t)\right]\flat^{\mu}\\ &\qquad\qquad\qquad\times\Theta(t+T/2)\Theta(T/2-t),
\end{split}\label{EField}
\end{equation}where $\flat^\mu=(0,0,1,0)$ is the polarization  four-vector and   $\Theta (x)$ denotes the unit step function: $\Theta (x)=1$ at $x\geqslant 0$, $\Theta (x)=0$ 
at $x<0$. Here the subscripts ``$s$'' and ``$w$'' are used to identify the strong and weak field strengths [$E_w\ll E_s$], respectively. 

Our investigation adopts the quantum  kinetic approach as theoretical tool to describe the production of electron-positron  pairs. This formulation--which is equivalent 
to other well-known approaches based on  QED in unstable vacuum \cite{Gitman,Fradkin}--comprises the  dynamical information of the  PP process in the single-particle distribution function  
$W(\pmb{p};t)$--summed over the  spin projections--of electrons and positrons to which the degrees of freedom in the external field are relaxed at asymptotically large times [$t\to\pm\infty$], 
i.e., when  the electric field is switched off $\pmb{E}(\pm\infty)\to 0$. The time evolution of this quantity is dictated by a quantum Boltzmann-Vlasov 
equation \cite{Schmidt:1998vi,Kluger,Schmidt:1999vi,alkofer:2001ib}, whose  integro-differential version:
\begin{equation}\label{vlasov}
\begin{split}
&\dot{W}(\pmb{p};t)=Q(\pmb{p},t)\int_{-\infty}^t d\tilde{t}\; Q(\pmb{p},\tilde{t})\\
&\qquad\qquad\times\left[1-W(\pmb{p};\tilde{t})\right]\cos\left[2\int_{\tilde{t}}^t dt^{\prime}\ \mathpzc{w}_{\pmb{p}}(t^{\prime})\right]
\end{split}
\end{equation}manifests both the nonequilibrium nature of the PP process and its non-Markovian feature.\footnote{The quantum field theoretical approach of the pair production problem--as 
encompassed by  Eq.~(\ref{vlasov}) concisely--can be formulated alternatively through a Riccati equation \cite{dumlu0,dumlu1} or via a representation involving three coupled ordinary differential 
equations. See for instance Refs.~\cite{Hebenstreit:2010vz,Otto:2014ssa,Panferov:2015yda}.}  The formula above assumes the vacuum initial condition $W(\pmb{p},-\infty)=0$ and applies the notation 
$\dot{W}(\pmb{p};t)\equiv\partial W(\pmb{p};t)/\partial t$. Besides, it is characterized by the function $Q(\pmb{p},t)\equiv e E(t) \epsilon_\perp /\mathpzc{w}_{\pmb{p}}^2(t)$,  
which depends on the transverse energy of the Dirac fermions $\epsilon_\perp=\sqrt{m^2+p_\perp^{\,2}}$ and the respective total energy squared
$\mathpzc{w}^2_{\pmb{p}}(t)=\epsilon_\perp^2+[p_\parallel-e\mathpzc{A}(t)]^2$ of an electron. Here $\pmb{p}_\perp=(p_x,0,p_z)$ and $\pmb{p}_\parallel=(0,p_y,0)$ 
are the components of the canonical momentum perpendicular and parallel to the direction of $\pmb{E}(t)$, respectively.

It is known that  Eq.~(\ref{vlasov}) can only be solved exactly for a few special backgrounds, e.g.,  constant and Sauter-type electric fields. Finding analytic 
solutions beyond the aforementioned  configurations is a difficult task. However, estimates can be obtained by using the low-density approximation [$W(\pmb{p};t)\ll 1$] within 
the Boltzmann-Vlasov equation. In such a case, the single-particle distribution function at time for which the field has been switched off  [$ W_T(\pmb{p})\equiv \lim_{t\to T}W(\pmb{p};t)$] 
can be approximated by \cite{Schmidt:1999vi,BlaschkeCPP}
\begin{equation}
 W_T(\pmb{p})\approx\frac{1}{2}\left\vert\int_{-\nicefrac{T}{2}}^{\nicefrac{T}{2}} d\tilde{t}\;Q_{\pmb{p}}(\tilde{t})e^{i\Lambda_{\pmb{p}}(\tilde{t})}\right\vert^2\label{lawdensitylimit}
\end{equation}with  $\Lambda_{\pmb{p}}(\tilde{t})\equiv\int_0^{\tilde{t}} d\mathpzc{t}\;\mathpzc{w}_{\pmb{p}}(\mathpzc{t})$. We note that the integration contained in this formula is nothing 
but the solution of the linearized Riccati equation  on which the study in Ref.~\cite{dumlu1} relies.  At this point it turns out to be rather  illuminating to  perform  the change of variables 
$\tau=[p_\parallel-e\mathpzc{A}(\tilde{t})]/\epsilon_\perp$ and $\tilde{\tau}=[p_\parallel-e\mathpzc{A}(\mathpzc{t})]/\epsilon_\perp$. As a consequence, the integral in Eq.~(\ref{lawdensitylimit}) 
becomes
\begin{equation}\label{intermediate1}
\begin{split}
&\int_{-\nicefrac{T}{2}}^{\nicefrac{T}{2}} d\tilde{t}\ldots=\int_{\frac{\gamma_\parallel}{\gamma_\perp}-\frac{\pi N}{\gamma_\perp}}^{\frac{\gamma_\parallel}{\gamma_\perp}+\frac{\pi N}{\gamma_\perp}} \frac{d\tau}{1+\tau^2}
\exp\left[\frac{\epsilon_\perp^2}{eE_s}\mathcal{S}(\tau)\right],\\
&\qquad\qquad\mathcal{S}(\tau)=2i\int_0^\tau d\tilde{\tau}\frac{(1+\tilde{\tau}^2)^{\nicefrac{1}{2}}}{1+\varepsilon \cos(\omega \mathpzc{t})},
\end{split}
\end{equation}where $\varepsilon=E_w/E_s\ll1$ parametrizes the relative weakness of the fast-oscillating  mode.  The expression above  constitutes the starting point for further considerations. 
In its second line, $\mathpzc{t}$ has to be considered as a function of $\tilde{\tau}$. However, this inversion cannot be determined analytically, but only through reversion of the corresponding 
series \cite{abramowitz,Kim2007}. In this case the leading order term 
\begin{equation}
\mathpzc{t}(\tilde{\tau})\approx\frac{1}{\omega}(\gamma_\perp\tilde{\tau}-\gamma_\parallel)
\end{equation}coincides with the inverse of the function $\tau(\mathpzc{t})$ averaged over a cycle of the weak field. Here, we have introduced the dimensionless parameters
\begin{equation}
\gamma_\parallel=\gamma\frac{p_\parallel}{m}\quad\mathrm{and}\quad\gamma_\perp=\gamma\frac{\epsilon_\perp}{m},
\label{inversionfull}
\end{equation}
Observe that, in the limit of $p_\perp\to 0$, $\gamma_\perp$ reduces to the 
combined Keldysh parameter [$\gamma=\omega m/(eE_s)$].  In order to suitably fit the external parameters to current and foreseeable experimental setups, we will suppose hereafter that  $m^2\gg(eE_s)$ 
and  $2m>\omega$. We note that  an assisted scenario with  $2m>\omega\gtrsim m$, is characterized by the restriction  $\gamma_\perp\geqslant\gamma\gg1$. Conversely, if $\omega\ll m^3/(eE_s)$--leading to  
$\gamma\ll1$--the effective reduction of the barrier width  between the negative and positive continuum is expected to be almost insignificant,  and the oscillating field  would not play a significant
role in the production of pairs.

\section{Properties of the particle spectrum \label{SPDFasymptotic}}

We wish to find closed-form analytic expressions for Eq.~(\ref{lawdensitylimit})  in the case characterized by the condition $\pi N\gg \gamma_\perp,\vert\gamma_\parallel\vert$. 
Therefore, the treatment developed in this subsection is  limited to small momentum components relative to the one associated with  the external field, i.e., to values  $\vert p_\parallel\vert \ll eE_sT/2$  
and  $\vert p_\perp\vert\ll eE_sT/2$. To facilitate the mathematical treatment of the problem, we will formally extend the outer integration limits in Eq.~(\ref{intermediate1}) 
to $\pm\infty$. As a consequence, the single-particle distribution function $W_T(\pmb{p})$ [see Eq.~(\ref{lawdensitylimit})] approaches to a $2\pi-$periodic function in 
$\gamma_\parallel$. Correspondingly, its dependence on this variable will be investigated in  the interval $-\pi<\gamma_\parallel\leqslant\pi$.

Since the  factor $\epsilon_\perp^2/(eE_s)\geqslant m^2/(eE_s)\gg1$, the exponential in Eq.~(\ref{intermediate1}) oscillates very fast and the steepest-decent method represents 
a suitable tool to carry out its outer integration. In order to apply this method, we first extend the integration variables to the complex $\tau-$plane. As it is characteristic 
in problems of this nature, the poles at  $\tau=\pm i$  are  also branch-points of the integrand. The branch-cuts are then chosen from $\tau=i$ to $\tau=i\infty$ and from $\tau=-i$ 
to $\tau=-i\infty$, i.e.,  $\tau^2+1=(\tau+i)(\tau-i)=\vert\tau+i\vert\vert\tau-i\vert e^{i\varphi_+}e^{i\varphi_{-}}$ with  $-3\pi/2\leqslant\varphi_+<\pi/2$  and $-\pi/2\leqslant\varphi_-<3\pi/2$ 
referring to the local polar angle linked to $\tau=+i$  and $\tau=-i$, respectively. Still, in the cut $\tau$-plane there exist poles linked to the integrand of  $\mathcal{S}(\tau)$: 
\begin{equation}\label{poles}
\tau_{\pm\mathpzc{k}}=\frac{\gamma_\parallel+(2\mathpzc{k}-1)\pi}{\gamma_\perp}\pm i\frac{\gamma_{\mathrm{cr}}}{\gamma_\perp},\qquad \mathpzc{k}\in \mathbb{Z},
\end{equation} where a loss of analyticity is exhibited. While the integer value $\mathpzc{k}$ manifests the  periodicity in $\gamma_\parallel$, the  critical Keldysh parameter \cite{Linder:2015vta,Togrimsson1,Togrimsson2,Akal:2018udh} 
\begin{equation}
\gamma_{\mathrm{cr}}=\ln\left(\frac{2}{\varepsilon}\right),\qquad \varepsilon\ll1\label{criticalgamma}
\end{equation}rules two different scenarios depending on whether $\gamma_\perp$ does or does not exceed the value of $\gamma_{\rm cr}$. Below  we will describe further this point.

\begin{figure}
\includegraphics[width=0.4\textwidth]{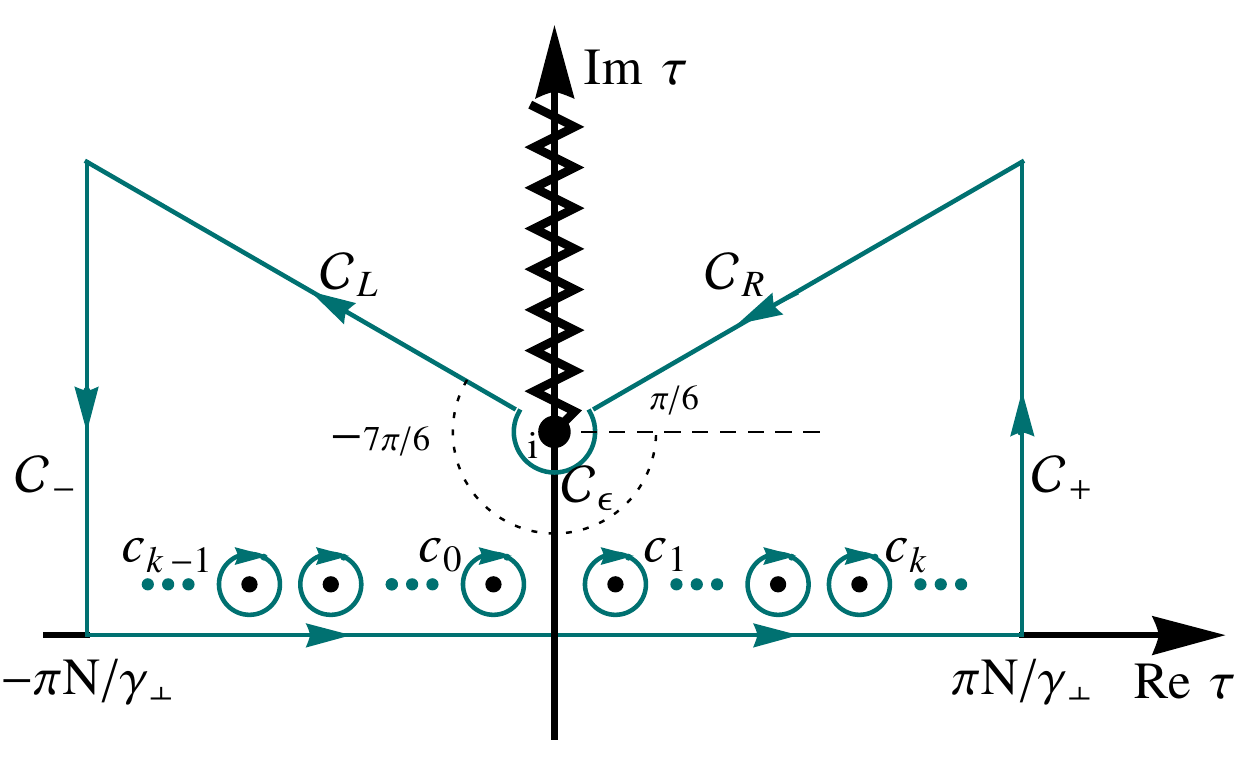}
\caption{Integration contour in the complex plane $\tau$ for $\arg\kappa\approx0$. The positive branch cut, defined from $z=i$ to $z=i\infty$, is depicted by a vertical 
zigzag line. Here the imaginary parts of the poles have been taken smaller than unity [$\gamma_{\rm cr}\ll\gamma_\perp$]. Shortcuts joining each circumventing circle  with 
the real $\tau-$axis have been omitted for simplicity. They give no contribution as, for each circle,  both lie infinitesimally close together and have opposite orientation.}
\label{fig:0}
\end{figure}

Now, the saddle-points linked to the exponent [$\tau=\pm i$] coincide with the branch-points found previously. In a vicinity of  $\tau=i$, the pre-exponential of the integrand behaves as 
$\sim 1/[2i(\tau-i)]$, whereas  
\begin{equation}
\begin{split}
&\mathcal{S}(\tau)\approx\frac{i}{2}\mathcal{S}_0+\frac{1}{\sqrt{2}}\vert \tau-i\vert^{\nicefrac{3}{2}}\left\vert\kappa\right\vert e^{i\arg\kappa+i\frac{3\pi}{4}+i\frac{3}{2}\varphi_+},\\
&\mathcal{S}_0=4\int_0^i d\tilde{\tau}\frac{(1+\tilde{\tau}^2)^{\nicefrac{1}{2}}}{1+\varepsilon\cos(\gamma_\perp\tau-\gamma_\parallel)}.
\end{split}
\label{behavior}
\end{equation}In the first line, $\arg\kappa$ denotes the principal value of the argument of  $\kappa=[1+\varepsilon\cos\left(\omega \mathpzc{t}(i)\right)]^{-1}$. Here, the directions of 
the steepest-descents can be locally approximated by choosing  $\varphi_+=-\frac{\pi}{2}+\frac{2}{3}k\pi-\frac{2}{3}\arg\kappa$ with $k\in \mathbb{Z}$  such that the condition 
$\cos(\arg\kappa+\frac{3}{4}\pi+\frac{3}{2}\varphi_+)<0$ holds. In connection  we find 
\begin{equation}\label{anglessd}
\begin{split}
\varphi_{+1}=\frac{\pi}{6}-&\frac{2}{3}\arg\kappa,\quad \varphi_{+2}=-\frac{7\pi}{6}-\frac{2}{3}\arg\kappa. 
\end{split}
\end{equation}Since the angles above  must lie  within the interval  $[-\frac{3\pi}{2},\frac{\pi}{2})$ [read  above Eq.~(\ref{behavior})], $\arg\kappa$ turns out to be restricted  to the region
$-\frac{\pi}{2}<\arg\kappa\leqslant\frac{\pi}{2}$ with
\begin{equation}
\begin{split}
&\arg\kappa=\tan^{-1}\left[\frac{\varepsilon \sinh(\gamma_\perp)\sin(\gamma_\parallel)}{1+\varepsilon \cosh(\gamma_\perp)\cos(\gamma_\parallel)}\right]+\mathpzc{s}_i
\end{split}\label{argumentphi}
\end{equation} with $\mathpzc{s}_{1,4}=0$, $\mathpzc{s}_{2}=\pi$, $\mathpzc{s}_{3}=-\pi$, where  the choice of the constant $\mathpzc{s}_{i}$ corresponds to suitable values in each  of the four 
quadrants \cite{Ablowitz}. Manifestly, Eq.~(\ref{argumentphi}) reveals that the directions of the steepest-descents change as  $\gamma_\perp$ and $\gamma_\parallel$ are varied. We remark that, 
in the absence of the perturbative mode [$\varepsilon=0\Leftrightarrow\arg\kappa=0$], these directions are fully specified: $\varphi_{+1}=\pi/6$ and $\varphi_{+2}=-7\pi/6$. It is worth remarking 
that these angles do not change appreciably when the fast oscillating field is turned-on [$\varepsilon\neq0$] and   $\gamma_{\rm cr}\gg\gamma_\perp$; no matter the value of $\gamma_\parallel$. 
The situation does not change either if $\varepsilon\neq0$, $\gamma_\perp\gg \gamma_{\rm cr}$ with $\gamma_\perp\gg1$ and $\vert\gamma_\parallel \vert\ll1$.  Even, for 
$\vert\gamma_\parallel\vert\ll1$ and $\gamma_\perp\sim\gamma_{\rm cr}>2 $ it can be verified that $\arg\kappa\approx 0$ and the directions of the steepest-descents approach locally to those 
arising when the strong electric field is present only. 

We take advantage of the described feature to deform the initial integration contour  [see Eq.~(\ref{intermediate1})] as depicted in Fig.~\ref{fig:0}. The portion covering 
the real axis deviates several times through circles $\mathpzc{c}_{\mathpzc{k}}$ with an infinitesimal radius [$\mathpzc{r}\to0$]. At this point it should be understood that each of them join the real 
$\tau-$axis via  parallel shortcuts with opposite directions. However, in the picture,  they  have been omitted for simplicity and because they do not contribute at all. Here $\mathpzc{k}$ 
labels a circumvented  pole between the most distant extremes of the circuit [see Eq.~(\ref{poles})]. In the picture, the imaginary parts of the poles have been taken with $\gamma_\perp\gg \gamma_{\rm cr}$. 
For $\gamma_\perp\ll \gamma_{\rm cr}$, the pole locations are moved upward, and many of them could lie above the path $\Gamma=\mathpzc{C_R}\cup\mathpzc{C_\epsilon}\cup\mathpzc{C_L}$. 
In such a situation, the contour of integration is chosen similar to the previous case: those poles remaining below $\Gamma$ are then eluded. Consequently, the  Cauchy's theorem  allows us to express  
\begin{equation}
\begin{split}
&\int_{-\infty}^\infty d\tau\ldots=-\exp\left [\frac{i \epsilon_\perp^2}{2 eE_s}\mathcal{S}_0\right]\\ &\qquad\quad\times \int_\Gamma \frac{d\tau}{2i(\tau-i)} \exp\left [\frac{\epsilon_\perp^2}{ \sqrt{2}eE_s} \kappa[i(\tau-i)]^{\nicefrac{3}{2}}\right],
\end{split}\label{sisi}
\end{equation}where we have taken into account that the contribution linked to each pole's  circumvention, i.e.,  over $\mathpzc{c}_{\mathpzc{k}}$, vanishes as  $\mathpzc{r}\to0$ [for details read 
the Appendix~\ref{AppendixA}]. Likewise,  we have considered that the  integrations over $\mathpzc{C_\pm}$  give no contributions when $N\to\infty$. The details of this considerations are summarized 
in  Appendix~\ref{AppendixB}. 

\begin{figure}
\includegraphics[width=0.4\textwidth]{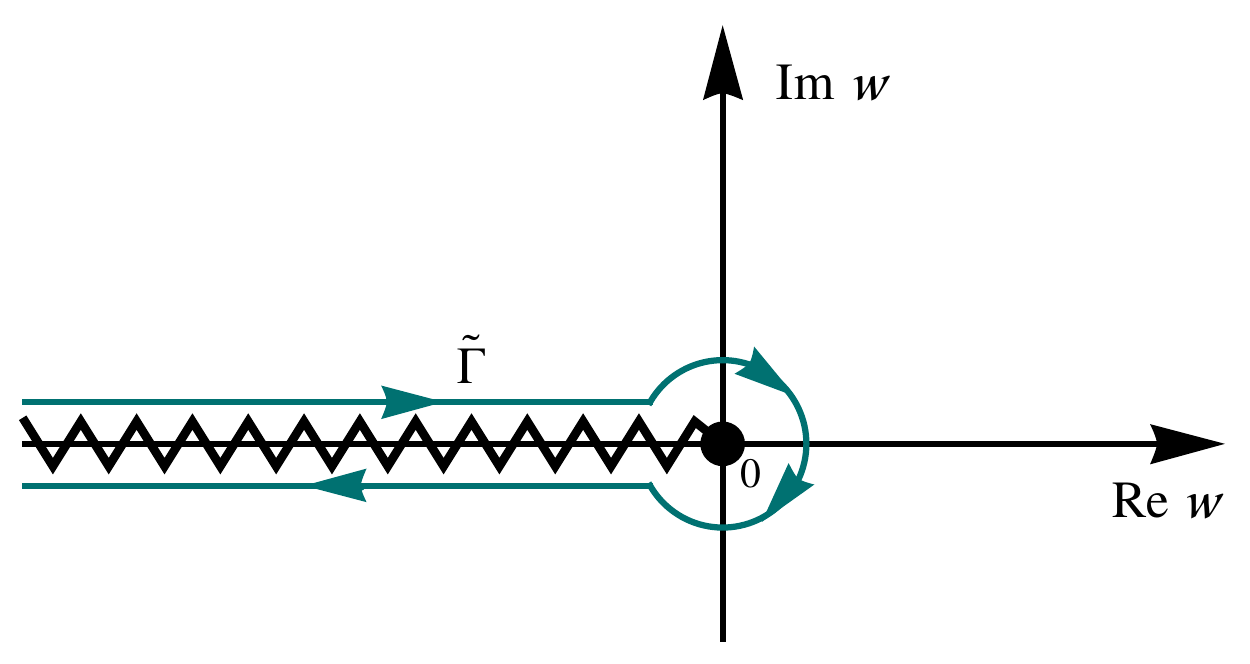}
\caption{Path  of integration in the complex $w-$plane. Note that it encloses the \emph{outer} region in counterclockwise sense.}
\label{fig:1}
\end{figure}

We point out  that  Eq.~(\ref{sisi}) can  also be applied to those cases in which the integration contour differs substantially from the one analyzed explicitly here,  provided  no contribution arises from those  circuits connecting the region 
$-\pi N/\gamma_\perp\leqslant\mathrm{Re}\;\tau\leqslant\pi N/\gamma_\perp$ with the sectors ending in the steepest descents. Hereafter we will suppose that this is the case. Thus, by 
performing the map $w=\epsilon_\perp^2\kappa[i(\tau-i)]^{\nicefrac{3}{2}}/[\sqrt{2}eE_s]$,  Eq.~(\ref{sisi})  becomes  
\begin{equation}\begin{split}
&\int_{-\infty}^\infty d\tau\ldots=\frac{2\pi i}{3}\exp\left [\frac{i \epsilon_\perp^2}{2 eE_s}\mathcal{S}_0\right]\oint_{\tilde{\Gamma}} \frac{dw}{2\pi w}e^w.
\end{split}\label{intermidiateintegrationforW}
\end{equation}Observe that  the closed path of integration $\tilde{\Gamma}$ [see  Fig.~\ref{fig:1}] encloses a single pole at infinity [$w=+\infty$]. Hence, the application of the residue theorem leads to
$\oint_{\tilde{\Gamma}} dw e^w/(2\pi w)=i$ \cite{Brezin}. We substitute this outcome into Eq.~(\ref{intermidiateintegrationforW}). The resulting expression is inserted  into Eq.~(\ref{lawdensitylimit}) 
afterwards. As a consequence, the single-particle distribution function reduces to
\begin{equation}
W_T(\pmb{p})\approx 2 e^{-\frac{\epsilon_\perp^2}{eE_s}\mathrm{Im}\;\mathcal{S}_0(\gamma_\perp,\gamma_\parallel)},
\label{distributionfunction}
\end{equation}where an unessential pre-exponential factor of the  order of unity has been omitted  [$\pi^2/9\approx1.1$]. We stress that  $\mathrm{Im}\;\mathcal{S}_0(\gamma_\perp,\gamma_\parallel)$ is given in 
Eq.~(\ref{behavior}). By taking $\tau=iy$, it reads 
\begin{equation}
\begin{split}\label{ImS}
&\mathrm{Im}\;\mathcal{S}_0(\gamma_\perp,\gamma_\parallel)=4\int_0^1dy\frac{\sqrt{1-y^2}\cos^2(\phi_y)}{1+\varepsilon\cosh(\gamma_\perp y)\cos(\gamma_\parallel)}.
\end{split}
\end{equation} Here the function $\phi_{y}$ is the angle determined by the $\tan^{-1}-$function involved in  Eq.~(\ref{argumentphi})  with $\gamma_\perp$ replaced by $\gamma_\perp y$. 
We remark that the expression above is an even function in $\gamma_\parallel$ which applies for any of the following two regimes:  $\gamma_{\mathrm{cr}}\gg\gamma_\perp$ and 
$\gamma_\perp\gg\gamma_{\mathrm{cr}}$ with $\gamma_\perp\gg1$. Hence  we further restrict our  investigation to  $0\leqslant\gamma_\parallel\leqslant\pi$.

Observe that, in the limit of $\varepsilon\to0$,  $\mathrm{Im}\;\mathcal{S}_0(\gamma_\perp,\gamma_\parallel)=\pi$. Hence, $W_T(\pmb{p})$ reduces to the known expression in a constant electric field \cite{Hebenstreit:2010vz}: 
$W_T(\pmb{p})\approx2\exp[-\pi\epsilon_\perp^2/(eE_{\mathrm{cr}})]$. The situation changes when $\varepsilon\neq0$. To show this, we investigate some analytic and asymptotic properties of Eq.~(\ref{ImS}). First 
of all, when setting  $\partial\mathrm{Im}\;\mathcal{S}_0(\gamma_\perp,\gamma_\parallel)/\partial\gamma_\parallel$ to zero, we find that the corresponding extreme points  are located at the borders of the region  
encompassed by  $0\leqslant\gamma_\parallel\leqslant\pi$.  To elucidate which of them maximize and minimize Eq.~(\ref{ImS}), we will establish its asymptotic formulas evaluated at $\gamma_\parallel=0$ and 
$\gamma_\parallel=\pi$, respectively. Let us begin with the case in which $\gamma_\perp\gg\gamma_{\rm cr}$ with $\gamma_\perp\gg1$. We then introduce a positive splitting  parameter $\mathpzc{y}_0$  satisfying the conditions $\gamma_\perp^{-1}\ll \mathpzc{y}_0\ll1$ and 
$\gamma_{\mathrm{cr}}/\gamma_\perp\ll \mathpzc{y}_0$. Afterward, the $y$ integration is divided   as follows:  $\mathrm{Im}\;\mathcal{S}_0(\gamma_\perp,\gamma_\parallel)=4\int_0^{\mathpzc{y}_0}dy\ldots+4\int_{\mathpzc{y}_0}^1dy\ldots$. In 
the integral defined over the region $[0,\mathpzc{y}_0]$ the integration  variable is very small [$y\ll1$], leading  to approximate the square root contained in the integrand by  unity [$(1-y^2)^{\nicefrac{1}{2}}\approx1$].  
Conversely, the main contribution to the integral defined over  $[\mathpzc{y}_0,1]$  results from those values of $y$ fulfilling the condition $y\gg\gamma_{\mathrm{cr}}/\gamma_\perp$,  in which case the  integrand can be 
approximated  by  $\sim \pm 2\exp[-\gamma_\perp y]/\varepsilon$. Consequently,  
 \begin{equation}
\begin{split}
\mathrm{Im}\;\mathcal{S}_0(\gamma_\perp,\gamma_\parallel)&\approx  4\int_0^{\infty}\frac{dy}{1\pm\varepsilon\cosh(\gamma_\perp y)}\\ &\pm\frac{8}{\varepsilon}\int_{\mathpzc{y}_0}^{1}dy e^{-\gamma_\perp y}\sqrt{1-y^2}\\ 
&- 4\int_{\mathpzc{y}_0}^{\infty}\frac{dy}{1\pm\varepsilon\cosh(\gamma_\perp y)}.
\end{split}
\end{equation} While the positive sign corresponds to $\gamma_\parallel=0$,  the negative one  is associated with $\gamma_\parallel=\pi$. Now, the first integral  involved in this expression  can be calculated by using Eq.~($3.513.2$) in Ref.~\cite{Gradshteyn}. The remaining two can be combined in an integral  independent of the parameter $\mathpzc{y}_0$.  
Indeed, note that  as $\gamma_\perp\gg1$, the integrand in $\int_{\mathpzc{y}_0}^{1}dy\ldots$ behaves as  $\sim \exp[\gamma_\perp y]$,  whereas the one  defined in $[\mathpzc{y}_0,\infty)$ can be approximated by  
$\sim \pm 2\exp[-\gamma_\perp y]/\varepsilon$.  Keeping all these details in mind we obtain  
\begin{equation}
\begin{split}
&\mathrm{Im}\;\mathcal{S}_0(\gamma_\perp,\gamma_\parallel)\approx 4 \left\{
\begin{array}{cc}
\displaystyle\frac{\gamma_{\mathrm{cr}}}{\gamma_\perp}-\frac{1}{\gamma_\perp}e^{-\gamma_\perp}& \mathrm{for}\  \gamma_\parallel=0,\\\\
\displaystyle\frac{\gamma_{\mathrm{cr}}}{\gamma_\perp}+\frac{1}{\gamma_\perp}e^{-\gamma_\perp}& \mathrm{for}\  \gamma_\parallel=\pi. 
\end{array}\right.
\end{split}\label{largekeldych}
\end{equation}

\begin{figure}
\includegraphics[width=0.45\textwidth]{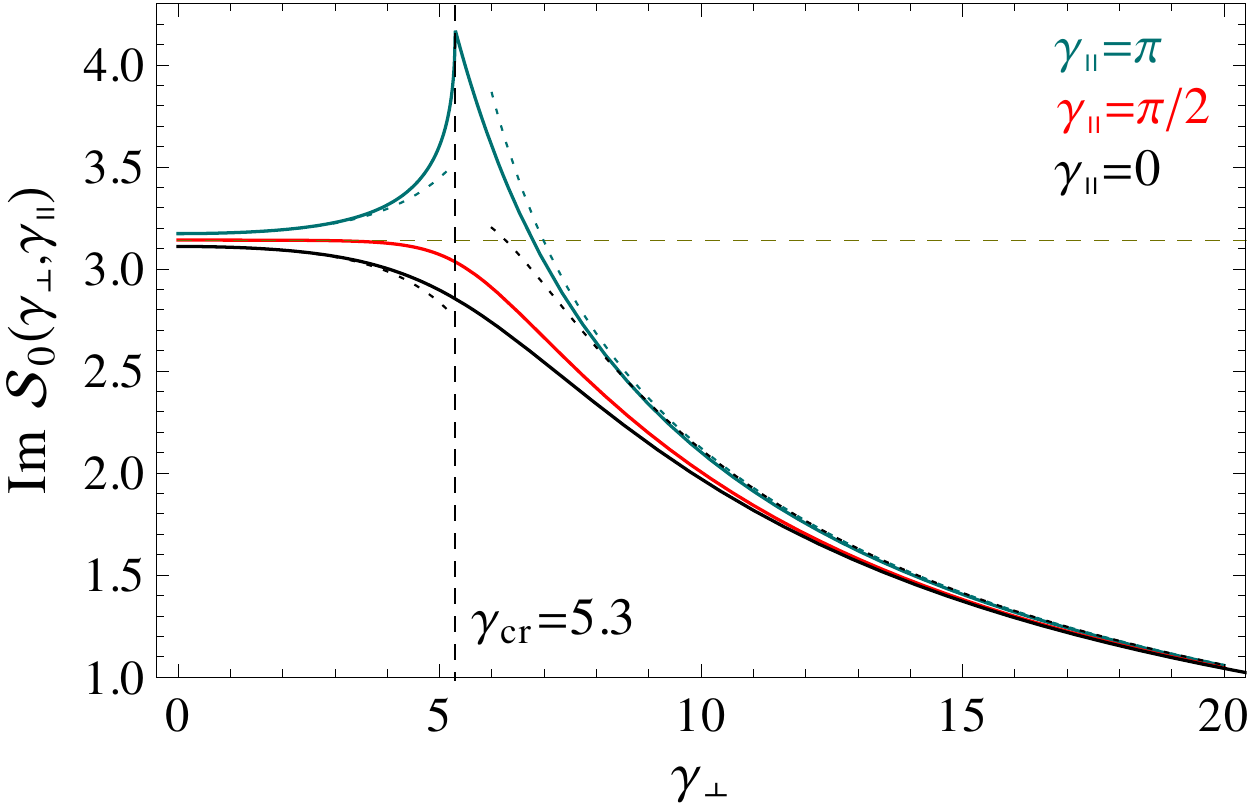}
\caption{Behavior of the function $\mathrm{Im}\;\mathcal{S}_0(\gamma_\perp,\gamma_\parallel)$ as a function of  $\gamma_\perp$: $\gamma_\parallel=0$ (black),  $\gamma_\parallel=\pi/2$ (red) and 
$\gamma_\parallel=\pi$ (darker cyan). The horizontal dashed line in olive corresponds to the value $\mathrm{Im}\;\mathcal{S}_0(\gamma_\perp,\gamma_\parallel)=\pi$ that results when the fast oscillating 
wave is not present, whereas the asymptotic trends [Eqs.~(\ref{smallkeldych}) and (\ref{largekeldych})] linked to the cases in which $\gamma_\parallel=0$ and $\gamma_\parallel=\pi$ are shown in dotted style. Here, 
the vertical dashed line indicates the critical value $\gamma_{\mathrm{cr}}\approx 5.3$ corresponding to $\varepsilon=10^{-2}$.  }
\label{fig:3}
\end{figure}

As the extreme points linked to $\gamma_\parallel$ coincide with the borders of the interval $[0,\pi]$, and $\mathrm{Im}\;\mathcal{S}_0(\gamma_\perp,\pi)>\mathrm{Im}\;\mathcal{S}_0(\gamma_\perp,0)$, the quantity  
$\mathrm{Im}\;\mathcal{S}_0(\gamma_\perp,\gamma_\parallel)$ as a function of $\gamma_\parallel$ grows monotonically in this region for any $\gamma_\perp$ satisfying the conditions   $\gamma_\perp\gg\gamma_{\rm cr}$ 
and $ \gamma_\perp\gg1$. This behavior  is also verified by looking at the solid curves in  Fig.~\ref{fig:3}, which have been obtained by setting $\gamma_\parallel=0$ (black), $\gamma_\parallel=\pi/2$ (red) and $\gamma_\parallel=\pi$ (darker cyan). 
Furthermore, the formula above reveals that the amplitude of oscillations in $\gamma_\parallel$  decreases as  $\sim 8 \exp[-\gamma_\perp]/\gamma_\perp$  when  $\gamma_\perp$ grows, a fact exhibited clearly in  Fig.~\ref{fig:3} 
[compare the solid curves in black and darker cyan at $\gamma_\perp=10$, for instance]. We remark that this trend changes as the condition  $\gamma_{\rm cr}\gg\gamma_\perp$ is fulfilled. Under this circumstance all instances in 
Eq.~(\ref{ImS}) depending  on $\varepsilon$ can be Taylor expanded. After the integration,
\begin{equation}
\begin{split}\label{smallkeldych}
\mathrm{Im}\;\mathcal{S}_0(\gamma_\perp,\gamma_\parallel)\approx\pi-\frac{2\pi}{\gamma_\perp}\varepsilon \mathpzc{I}_1(\gamma_\perp)\cos(\gamma_\parallel),
\end{split}
\end{equation}where Eq.~($3.534.1$) in Ref.~\cite{Gradshteyn} has been used. Here,  $\mathpzc{I}_1(x)=\frac{x}{2}+\frac{x^3}{2^24}+\frac{x^5}{2^24^26}+\ldots$ denotes the modified Bessel function of the first kind  with  
order one \cite{NIST}. The formula above coincides with Eq.~(37) in  Ref.~\cite{Linder:2015vta}. It  manifests clearly  a monotonic growing in $0\leqslant\gamma_\parallel\leqslant\pi$.
Furthermore, in the current regime, the amplitude of oscillation in $\gamma_\parallel$ scales as  $\sim 2\pi\varepsilon$.

Interestingly, when $\gamma_\perp$ moves away from $\gamma_{\rm cr}$ toward  larger values, all solid curves in Fig.~\ref{fig:3} show a significant falling as compared to  the characteristic value linked to the constant 
field case  [$\mathrm{Im}\;\mathcal{S}_0\approx\pi$].  Indeed, at $\gamma_\perp=20$, they have fallen to $\mathrm{Im}\;\mathcal{S}_0\approx1$.  
Besides, as Eq.~(\ref{largekeldych}) applies for $\gamma_\perp\gg \gamma_{\mathrm{cr}}$ with $\gamma_\perp\gg1$, the exponential contributions can be safely ignored. This fact implying that $\mathrm{Im}\;\mathcal{S}_0(\gamma_\perp,\gamma_\parallel)$ 
does not undergo an appreciable variation in  $\gamma_\parallel$. Keeping all these details in mind, we find that the single-particle distribution function  [see Eq.~(\ref{distributionfunction})] behaves as   
\begin{equation}
W_T(\pmb{p})\approx2\left(\frac{\varepsilon}{2}\right)^{\frac{4\epsilon_\perp}{\omega}}, \qquad\gamma_\perp\gg\gamma_{\mathrm{cr}},\quad \gamma_\perp\gg1.
\label{asymptoticforW}
\end{equation} This expression deserves further comments. Firstly, at $\pmb{p}=0$, the exponent associated with this asymptotic formula coincides with the minimal ``number'' of quanta  necessary to produce a pair at rest from 
the weak mode solely. This observation already provides evidences that an enhacement in $W_T(\pmb{p})$ could take place via the absorption of a quantum from the fast-oscillating field as compared with the case in 
wich only a constant electric field drives the vacuum instability. Indeed, let us suppose the particles are created in an assisted field setup characterized by the following parameters: $\omega= 1.7\; m$, $E_s=10^{-1}\; E_{\mathrm{cr}}$, 
i.e. $\gamma\approx17$, and $E_w=10^{-3}\;E_{\mathrm{cr}}$ corresponding to $\varepsilon=10^{-2}$ [$\gamma_{\rm cr}\approx5.3$].  Under such circumstances we find that $W_T(\pmb{0})\sim 10^{-5}$ exceeds by  $9$ orders 
of magnitude  the corresponding distribution function $W_T(\pmb{0})\sim 10^{-14}$ of the standard Schwinger mechanism. Apparently the enhancement becomes stronger as $E_s$ decrease gradually. However, it is worth 
pointing out that such a trend  is justified whenever $W_T(\pmb{p})$ remains smaller than unity [read discussion above Eq.~(\ref{lawdensitylimit})]. Hence, at $\pmb{p}=\pmb{0}$, this condition translates into the 
restriction $E_s\gg \frac{1}{2}E_w 2^{\frac{\omega}{4m}}$,  and by using the parameters above, for instance, it will imply that  $E_s\gg7\times 10^{-4} E_{\rm cr}$. 

\section{Pair production rate \label{observabledcp}}

The density of created  electron-positron pairs follows from integration of the  single-particle distribution function over the three momentum components. 
When exploiting both the cylindrical symmetry of our problem and the   $2\pi m\gamma^{-1}-$periodicity  of $W_T(\pmb{p})$  in $p_\parallel$  [see below Eq.~(\ref{ImS})] we can write 
\begin{equation}
\begin{split}
\mathpzc{N}&=\int \frac{d^3p}{(2\pi)^3}\; W_T(\pmb{p}) \\
&=\frac{eE_sT}{2\pi^2}\int_{0}^\pi d\gamma_\parallel\int_0^\infty\frac{dp_\perp}{2\pi}\;p_\perp W_T(\pmb{p}),
\end{split}\label{Ninitial}
\end{equation}where the even feature of $W_T(\pmb{p})$ in $p_\parallel$ has also been used. To satisfy the condition under which $W_T(\pmb{p})$ was derived [see first paragraph in Sec.~\ref{SPDFasymptotic}], the integral over $p_\perp$ must be performed 
over the region $[0,eE_sT]$. However,  the fast  damping  of its integrand in this variable allows us  to  extend  its  upper integration limit to infinity with no appreciable error.  

At this point,  it turns out to be convenient to  carry out  the change of variable $s^2=\omega^2\epsilon_\perp^2/(eE_s)^{2}$ in Eq~(\ref{Ninitial}) and to go over to the rate of created pairs. Consequently,  
\begin{equation}
\begin{split}
&\dot{\mathpzc{N}}\equiv\frac{\mathpzc{N}}{T}\approx\frac{(eE_s)^{3}}{2\pi^3\omega^2}\int_0^{\pi}d\gamma_\parallel \int_\gamma^\infty ds\\ 
&\qquad\qquad\quad\times s \exp\left[-\frac{eE_s}{\omega^2}s^2\mathrm{Im}\;\mathcal{S}_0(s,\gamma_\parallel)\right],
\end{split}\label{dfrt}
\end{equation}where Eq.~(\ref{distributionfunction}) has been inserted. Observe that the function $s^2\mathrm{Im}\mathcal{S}_0(s,\gamma_\parallel)$ grows monotonically in both  $s$ and $\gamma_\parallel$ and 
$\frac{eE_s}{\omega^2}s^2\mathrm{Im}\;\mathcal{S}_0(s,\gamma_\parallel)\geqslant \frac{E_{\mathrm{cr}}}{E_s}\mathrm{Im}\;\mathcal{S}_0(\gamma,\gamma_\parallel)\gg1$.\footnote{We note that the insertion of 
Eq.~(\ref{largekeldych}) into the condition $\frac{E_{\mathrm{cr}}}{E_s}\mathrm{Im}\;\mathcal{S}_0(\gamma,\gamma_\parallel)\gg1$ provides a restriction $E_s\gg \frac{1}{2}E_we^{\frac{\omega}{4m}}$, which is similar to the one 
given at the end of Sec.~\ref{SPDFasymptotic}.}  Therefore, we integrate by parts in $s$ and expand the resulting integrand around  $\gamma_\parallel\sim 0$.  Consequently,
\begin{equation}
\begin{split}
&\dot{\mathpzc{N}}\approx\frac{(eE_s)^{2}}{8\pi^4}\frac{\mathrm{erf}\left(\sqrt{\frac{1}{2}\frac{m}{\omega}\mathpzc{h}_\varepsilon(\gamma)}\pi\right)}{1-\mathpzc{g}_\varepsilon(\gamma)-\frac{1}{2}\gamma \mathpzc{g}_\varepsilon^\prime(\gamma)}\\
&\qquad\qquad\times\sqrt{\frac{2\pi\omega}{m \mathpzc{h}_\varepsilon(\gamma)}}\exp\left[-\pi\frac{m^2}{eE_s}\left\{1-\mathpzc{g}_\varepsilon(\gamma)\right\}\right]
\end{split}\label{main}
\end{equation}with $\mathrm{erf}(x)=\frac{2}{\sqrt{\pi}}\int_0^xdte^{-t^{2}}$ denoting the error function \cite{NIST}. The formula above  constitutes our main analytic result. It allows us in particular to obtain the scaling 
behavior of the process with the field parameters, and thus,  offers genuine advantages for optimizing the pair production yield and discriminating it from undesirable backgrounds. 
\begin{figure}
\includegraphics[width=0.45\textwidth]{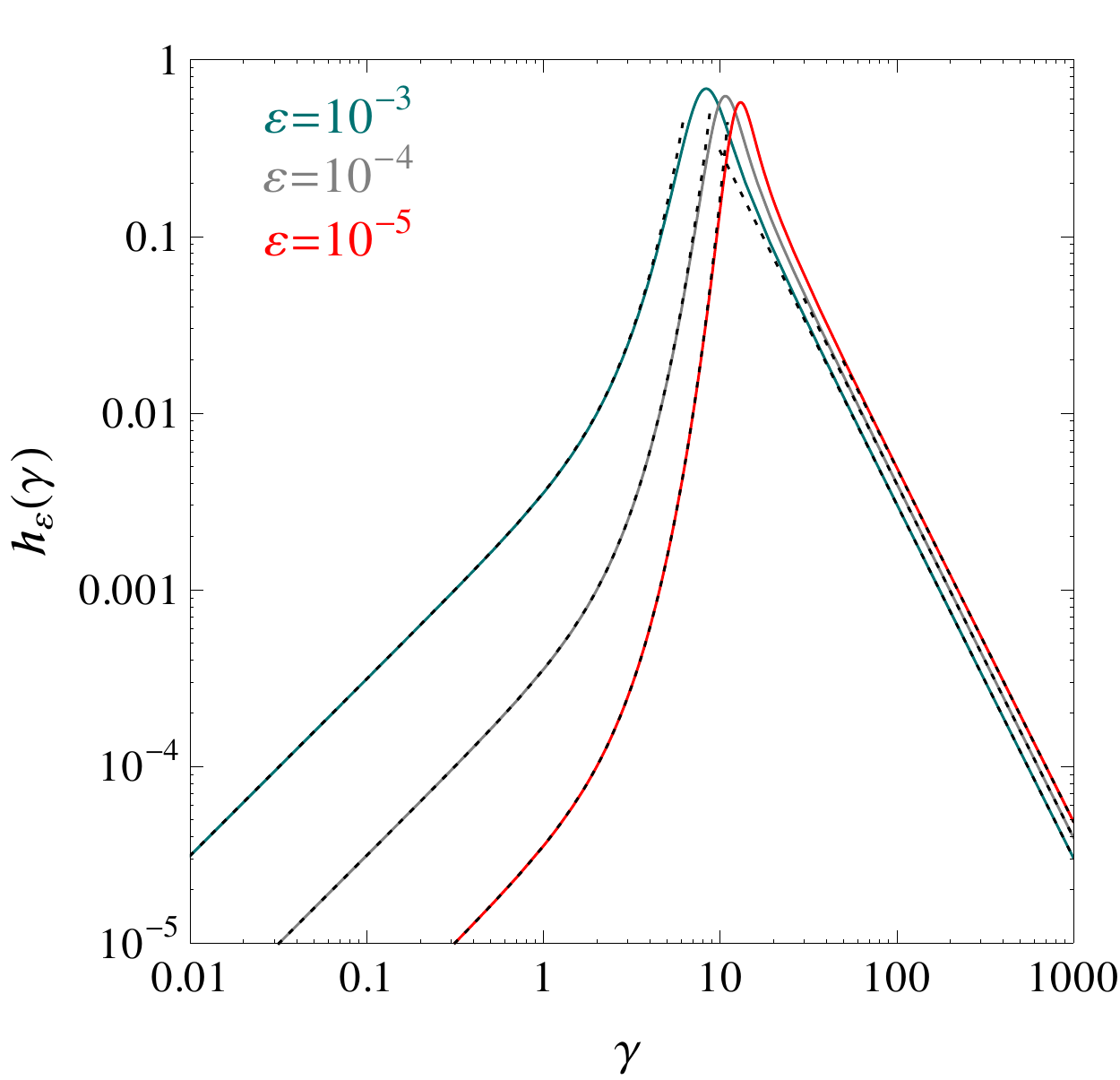}
\caption{Behavior of the function $\mathpzc{h}_\varepsilon(\gamma)$ [see Eq.~(\ref{functionh})].   The asymptotic trends [Eqs.~(\ref{asymptich})] linked to each curve are shown in dotted style.}
\label{fig:extfuf}
\end{figure}
\begin{figure*}
\includegraphics[width=0.45\textwidth]{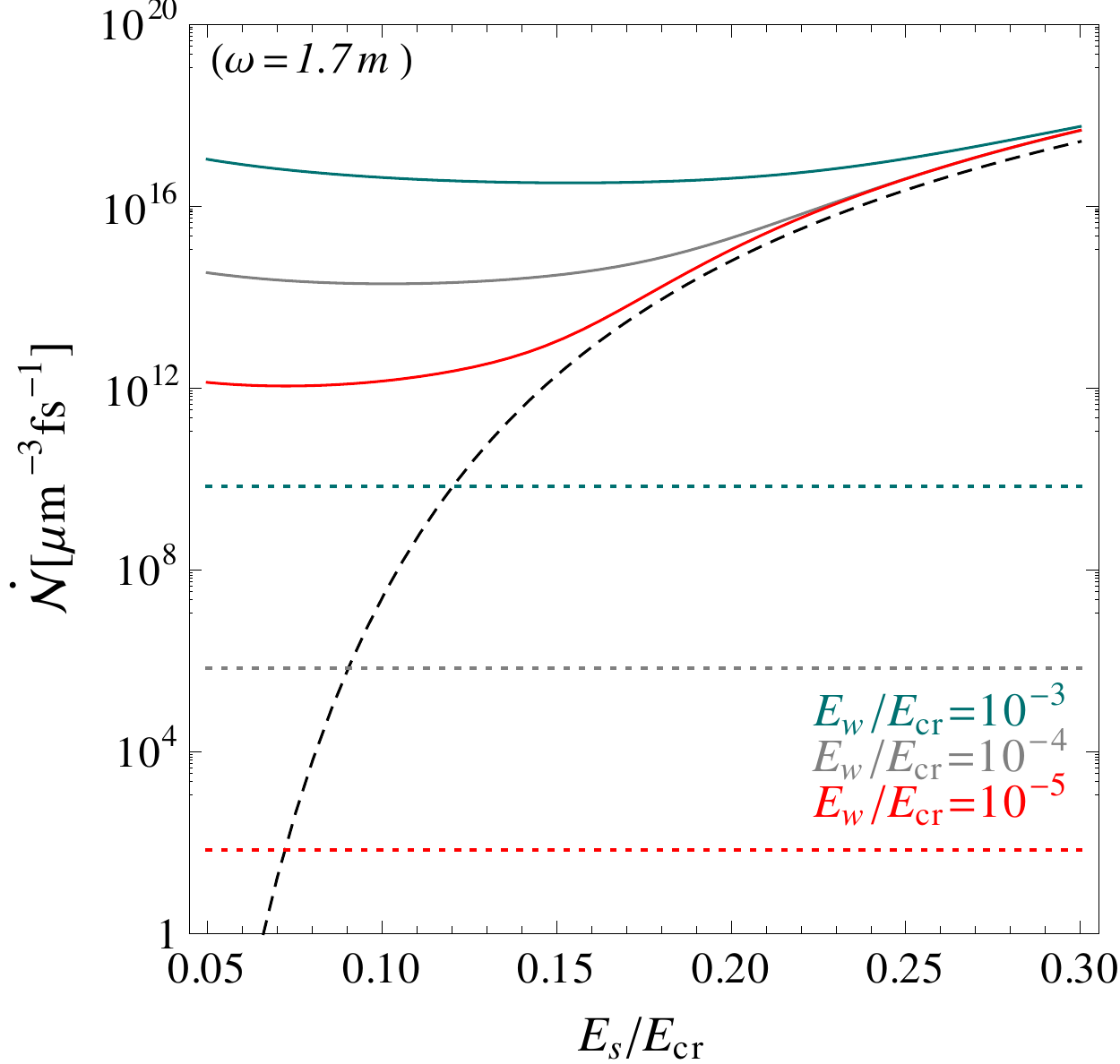}
\includegraphics[width=0.463\textwidth]{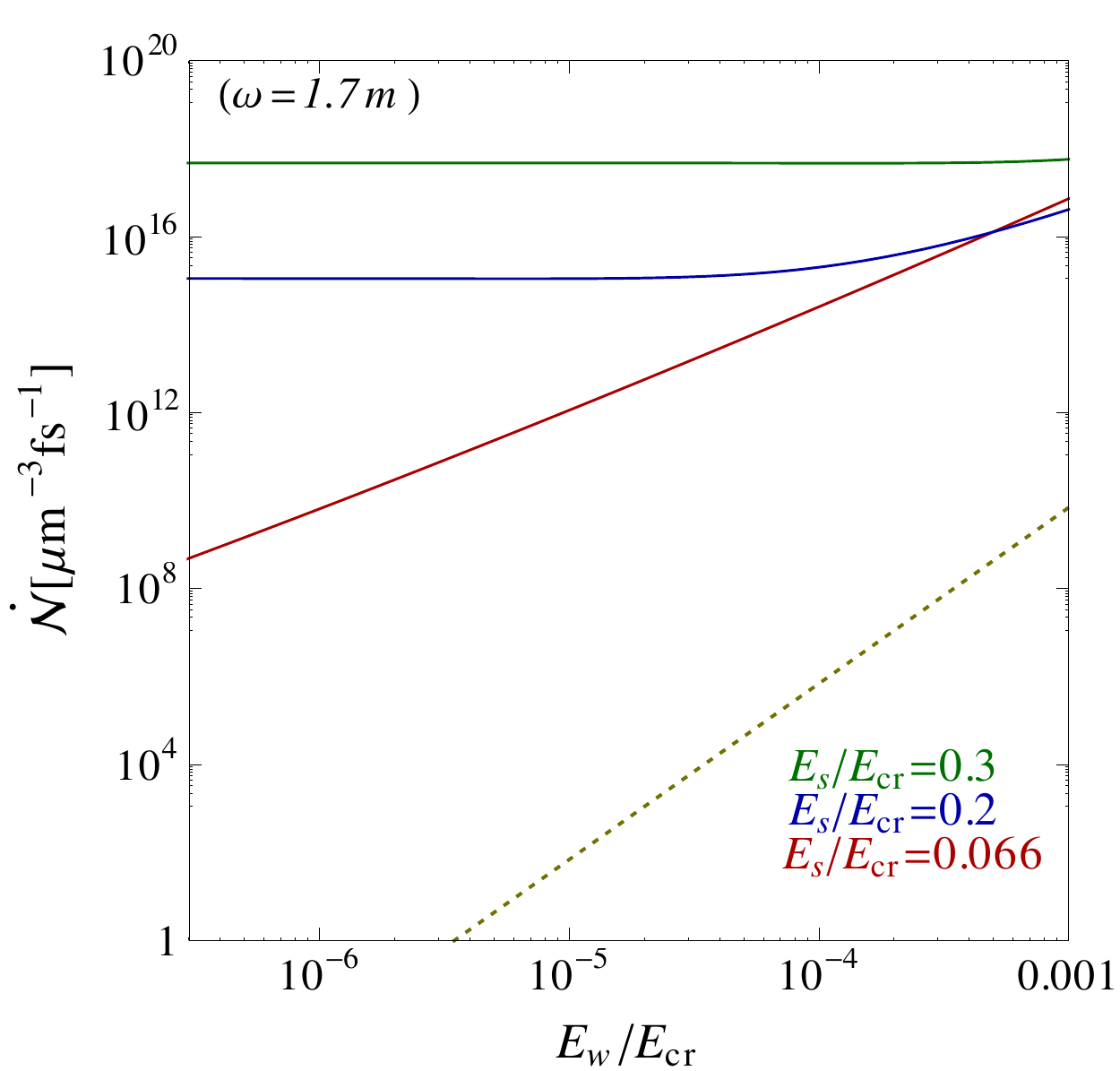}
\caption{Pair production rate per unit of volume  in a field configuration in which a constant  strong field is assisted by a weak but fast-oscillating electric mode. Its dependences on the strong and weak fields 
are shown in the left and right panel, respectively. In the left panel, the  dashed curve  results from the  standard Schwinger mechanism, whereas the horizontal dotted lines have been obtained by considering the effect 
of the fast-oscillating field only. Likewise, in the right panel, the dotted curve in olive describes the behavior of the rate when the strong field is not present and only the perturbative mode drives the pair creation 
process. The corresponding expression associated with this scenario can be found in Ref.~\cite{popov,Popov2002,Ringwald}.}
\label{fig:4}
\end{figure*}
Notice that its pre-exponential portion  
contains the functions $\mathpzc{g}_\varepsilon(\gamma)$  and $\mathpzc{g}_\varepsilon^\prime(\gamma)\equiv \left.\partial \mathpzc{g}_\varepsilon(s)/\partial s \right\vert_{s=\gamma}$  with
\begin{equation}
 \begin{split}
  \mathpzc{g}_\varepsilon(\gamma)&=1-\frac{1}{\pi}\mathrm{Im}\;\mathcal{S}_0(\gamma,0)\\
 &\approx\left\{\begin{array}{ccc}\displaystyle
 \frac{2\varepsilon}{\gamma}I_1(\gamma)&\mathrm{for}& \gamma\ll1,  \\ \displaystyle
 1-\frac{4}{\pi}\frac{\gamma_{\mathrm{cr}}}{\gamma}&\mathrm{for}&
 \begin{array}{c}
 \gamma\gg1,\\ \;\; \gamma\gg\gamma_{\mathrm{cr}}
 \end{array}
 \end{array}\right.
 \end{split}
\end{equation}
characterizing the decrement of the exponential function. As before, $\mathpzc{I}_1(x)=\frac{x}{2}+\frac{x^3}{2^24}+\frac{x^5}{2^24^26}+\ldots$ refers to the modified Bessel function of the 
first kind  with  order one  \cite{NIST}. Additionally, Eq.~(\ref{main}) introduces the function
\begin{equation}
\begin{split}
\mathpzc{h}_\varepsilon(\gamma)&\equiv\gamma\left.\frac{\partial^2}{\partial\gamma_\parallel^2}\; \mathrm{Im}\;\mathcal{S}_0(\gamma,\gamma_\parallel)\right\vert_{\gamma_\parallel=0}\\
&=8\varepsilon\gamma\int_0^1dy\frac{\sqrt{1-y^2}}{[1+\varepsilon\cosh(\gamma y)]^3}\\
&\quad\times\left[\frac{1}{2}\cosh(\gamma y)+\varepsilon\left(1-\frac{1}{2}\cosh^2(\gamma y)\right)\right],
 \end{split}\label{functionh}
\end{equation} whose behavior as a function of the parameter $\gamma$ is depicted in Fig.~\ref{fig:extfuf}. All curves exhibited there remain below one [$\mathpzc{h}_\varepsilon(\gamma)<1$]. 
They manifest fast decaying laws for both $\gamma\gg1$ and $\gamma\ll1$. These trends can be understood when looking for their asymptotes, which turn out to be:
\begin{equation}\label{asymptich}
 \mathpzc{h}_\varepsilon(\gamma)\approx\left\{
 \begin{array}{ccc}\displaystyle
  2\pi\varepsilon I_1(\gamma)&\mathrm{for}& \gamma\ll1, \\ 
 \displaystyle
 4\frac{\gamma_{\mathrm{cr}}}{\gamma^2}&\mathrm{for}& \begin{array}{c}
 \gamma\gg1,\\ \;\; \gamma\gg\gamma_{\mathrm{cr}}
 \end{array}.
 \end{array}\right.
\end{equation}We note that the expression linked to the case $\gamma\ll1$  can even be applied to study the behavior of  $\mathpzc{h}_\varepsilon(\gamma)$ in regions for which 
$\gamma\gtrsim1$.  The  dotted curves in the left portion of Fig.~\ref{fig:extfuf}  reveal us that the loss of accuracy in such a case is almost undiscernible.

In the regime characterized by the conditions $\gamma\ll 1$ and $m<\omega\lesssim2 m$, the argument of the error function is very small and its leading 
order contribution behaves as $\mathrm{erf}(x)\sim 2x/\sqrt{\pi}$. In such a scenario, Eq.~(\ref{main}) reproduces  the known rate for the Schwinger mechanism [quasi-static limit]
\begin{equation}
\begin{split}
&\dot{\mathpzc{N}}\approx\frac{(eE_s)^2}{4\pi^3} \exp\left[-\pi\frac{E_{\mathrm{cr}}}{E_s}\right].
\end{split}\label{schwingerrate}
\end{equation}When $\gamma$ exceeds unity and the critical value $\gamma_{\mathrm{cr}}$  significantly [$\gamma\gg1$ and $\gamma\gg \gamma_{\mathrm{cr}}$] with $m<\omega\lesssim2 m$, the small argument behavior of the error function still can be applied to 
Eq.~(\ref{main}) and the production rate  approximates  
\begin{equation}
\begin{split}
&\dot{\mathpzc{N}}\approx\frac{eE_s}{8\pi^2}\frac{m\omega}{\ln\left[2\frac{ E_s}{E_w}\right]}\left[\frac{1}{2}\frac{E_w}{E_s}\right]^{\frac{4m}{\omega}}.
\end{split}\label{assistedrate}
\end{equation}Interestingly, this formula  manifests a nonperturbative dependence not only in $E_s$  but also in the field strength  $E_w$  associated with the pertubative electric mode. Clearly,  in the intermediate regime 
not covered by these asymptotic cases, the  density rate for pair production  [see Eq.~(\ref{main})]  mixes  both   tunneling and multi-photon effects.  

\begin{figure}
\includegraphics[width=0.45\textwidth]{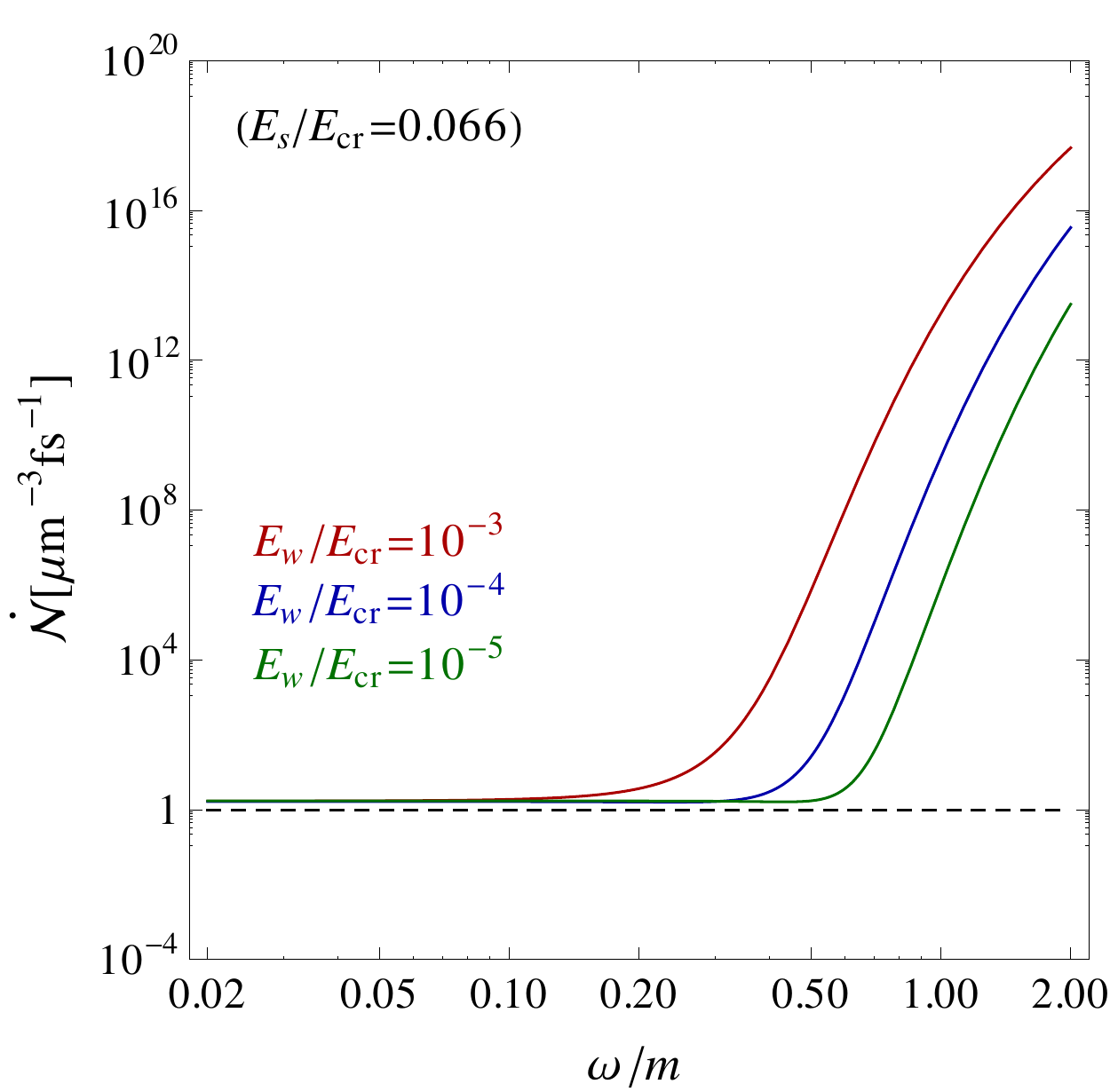}
\caption{Frequency dependence of the pair production rate per unit of volume associated with the  assisted Schwinger mechanism. Here the horizontal dashed line follows from the expression associated with the 
Schwinger mechanism.}
\label{fig:5}
\end{figure}

In order to extend further our knowledge on $\dot{\mathpzc{N}}$, we show in  Fig.~\ref{fig:4}  its behavior as a function of the strong [left panel] and weak [right panel] field strengths. The results exhibited in both 
panels  have been obtained by setting the frequency of the fast-oscillating field to $\omega=1.7 m$. Following the discussion at the end of Sec.~\ref{SPDFasymptotic}, the left  panel has been generated by varying the 
strong field  betwen $5\times 10^{-2}\leqslant E_s/ E_{\rm cr}\leqslant 3\times 10^{-1}$. Here the dashed curve corresponds to the  standard Schwinger mechanism [see Eq.~(\ref{schwingerrate})], whereas the dotted curves 
result from the case in which the pair production process is driven by the fast-oscillating mode only. The expression used to generate the curves linked to the latter  scenario is given in Refs.~\cite{popov,Popov2002,Ringwald}. 
Here, by increasing the weak field amplitude by an order of magnitude, the production rate grows by a factor $10^4$, indicating that the process occurs in the perturbative regime with absorption of two quanta $\omega$. 
For comparison we note that the curves for the assisted setup differ by relative factors of about $200$ each at   $E_s\approx 6.6\times 10^{-2} E_{\rm cr}$. We observe besides that, in average, the slope linked to the rate of the standard 
Schwinger effect, is larger than the ones corresponding to the assisted setup. Hence,  as the ratio $E_s/ E_{\rm cr}$ grows the enhancing due to the fast-oscillating field becomes less pronounced. This fact corroborates 
the  idea that, in an assisted scenario, there exist  two channels for increasing the pair production rate, either by growing the strong  field strength or via the absorption of  quanta from the fast-oscillating mode. The 
former path rules the process as $E_s$ grows,  while the latter dominates as the contrary condition occurs. 

Fig.~\ref{fig:5} is intended to provide insight about the trend of the rate with the change  of $\omega$. It has been obtained by setting the strong  field  to $E_s= 6.6\times 10^{-2}E_{\rm cr}$. As in 
Fig.~\ref{fig:4}, the dashed line follows from the expression associated with the Schwinger mechanism [see Eq.~(\ref{schwingerrate})]. We note that all solid curves closely approach to the value of the standard Schwinger 
mechanism when $\omega\ll m$.

We wish to put our outcomes in context. So,  let us suppose an experiment driven by the strong field to be reached at the forthcoming  ELI laser system.  Accordingly, we take  $E_s\approx 6.6\times 10^{-2} E_{\rm cr}$ 
as a reference parameter.  To achieve a peak field strength of this nature a strong focusing--close to the diffraction limit--is required. Consequently, we will suppose that both the spatial extent and the temporal 
length of the laser pulse are $\ell\sim 1\; \rm \mu m$ and  $T\sim 1\; \rm fs$, respectively. This laser is expected to operate with a central frequency $\Omega\approx1.55\;\rm eV$. Notice that, if the fast-oscillating 
mode operates at $\omega\approx1.7 m$, the number of cycles it makes during the pulse length $T$ is  $N\approx2.1\times 10^5$  which exceeds largely the combined Keldysh parameter $\gamma\approx26$.  Under such a circumstance 
the number of pairs yielded from an assisted configuration characteristized by  $E_w=10^{-3}E_{\rm cr}$ is $\mathpzc{n}_{e^-e^+}=\dot{\mathpzc{N}}\ell^3T\approx7.4\times 10^{16}$ [curve in cyan in the left panel of Fig.~\ref{fig:4}] would exceed the result associated with the  Schwinger mechanism [black dashed curve] 
$\mathpzc{n}_{e^-e^+}\sim 1$  by sixteen orders of magnitude roughly. However, when comparing this value with the dottet curve in cyan--corresponding to the case in which PP is due to the fast-oscillating mode only--we find 
$\mathpzc{n}_{e^-e^+}\approx7\times 10^{9}$, which leads to an effective enhancement by seven orders of magnitude, approximately. 

Some comments are in order. Firstly, we note that the discussed improvements have been obtained from  a single-laser shot only.  Certainly, our study provides evidences that the pair production enhancement increases 
significantly for frequencies $m\lesssim\omega<2m$. We note that, in such an energy regime, differences between pair  production in an oscillating electric field and pair production in a standing laser wave have been 
pointed out, owing to the spatial dependence and magnetic component of the latter \cite{Ruf,Dresden,Grobe3}. Hence, the results presented here are expected to describe only qualitatively the phenomenon taking  place in 
realistic laser fields. Notwithstanding this, the general physical conclusions drawn regarding the dependence on the electric field strengths of both the strong and the  fast-oscillating fields are expected to find their 
counterparts in a laser-based experiment for the assisted Schwinger mechanism.


\section{High-energy photon  emission  as a probe of the dynamically-assisted Schwinger mechanism \label{emissionsec}}

\begin{figure}
\includegraphics[width=3.45 in]{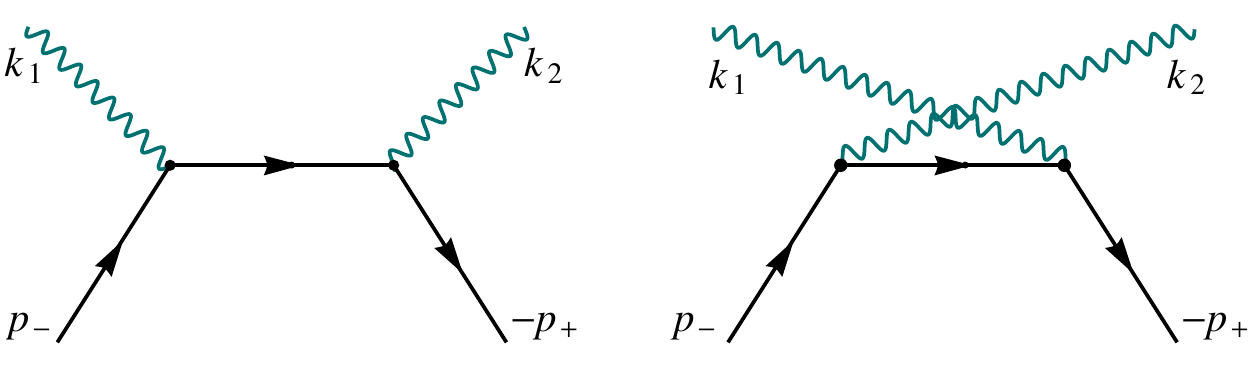}
\caption{Feynman diagrams contributing to the electron-positron recombination into two photons.}
\label{fig:ext}
\end{figure}

The  enhacement  induced  by  the superposition of the fast-oscillating  wave onto the strong field background might  facilitate the experimental verification of  the  spontaneous production of pairs 
from the vacuum, by detecting the generated particles directly.  However, the high number densities of created pairs promote  their recombinations [see  Fig.~\ref{fig:ext}],  and thus, the emission of photons.  
Hence, the created electron-positron plasma has a pronounced unstable  nature and the number of pairs evolves according to the law $\mathpzc{n}_{e^-e^+}(t)=\mathpzc{n}_{e^-e^+}[1+\delta t/\tau]^{-1}$,\footnote{This 
decay law arises when assuming that the associated rate is proportional to both the number of  electrons and of positrons present at time $t$, i.e. for $\dot{\mathpzc{n}}_{e^-e^+}(t)=-\frac{1}{\tau}\mathpzc{n}_{e^-e^+}^2(t)$.} 
where $\mathpzc{n}_{e^-e^+}$ is the number of pairs at the moment when the external field is switched off [$\delta t\equiv t-T/2=0$], whereas  $\tau$ denotes the plasma life time. An estimate for $\tau$ will 
be given below. It is worth remarking that the number of photons emitted  simultaneously at  time $t$ is  
\begin{equation}
\begin{split}
\Delta\mathpzc{n}_{\gamma}(t)&=\mathpzc{n}_{\mathrm{total}}(t)-\mathpzc{n}_\gamma(T)\\
&=\mathpzc{n}_{e^-e^+}\frac{\delta t}{\tau}\frac{1}{1+\frac{\delta t}{\tau}}.
\end{split}
\end{equation}While $\mathpzc{n}_{\mathrm{total}}(t)$ denotes the total number of photons present at the time $t$,  $\mathpzc{n}_\gamma(T)$ refers to the  existing amount of them  when the external field is 
switched off.\footnote{The  pair annihilation  occurring during the time in which the field is on has been analyzed numerically  based on  generalized  Vlasov equations \cite{Ruffini}. Likewise, the electron-positron 
recombination driven by a plane-wave  has been investigated in Refs.~\cite{Henrich,Ilderton}.} In  what follows,  we  suppose the latter causes only a minor  effect  on the single-particle distribution 
function $W_T(\pmb{p})$  [see Eqs.~(\ref{distributionfunction}), (\ref{smallkeldych}) and (\ref{asymptoticforW})] and  limit ourselves to the number of photons produced from the moment on when the field is switched off.

Clearly, in a  time interval  of the order of  $\tau$ or  larger--in addition to the electron-positron annihilation process--also scattering events of particles and antiparticles are very  likely to take 
place. As a consequence, the initial particle spectrum is supposed to change significantly.  In contrast,  for a $\delta t\ll \tau$, i.e. for early times, neither the single-particle distribution function nor the initial number 
of pairs are expected to change appreciably [$\mathpzc{n}_{e^-e^+}(t)\approx\mathpzc{n}_{e^-e^+}$]. Likewise, the total number of emitted photons approaches 
\begin{equation}
\Delta\mathpzc{n}_{\gamma}(t)\approx \mathpzc{n}_{e^-e^+}\frac{\delta t}{\tau}. \label{earlytimes}
\end{equation}
This formula constitutes  a good approximation only when  $\Delta\mathpzc{n}_{\gamma}(t)\ll  \mathpzc{n}_{e^-e^+}$, in which case the consequences of both annihilation and scattering processes  can be 
treated perturbatively. Under such circumstances,  the differential number of  recombination events  per unit volume and unit time approaches:
\begin{equation}
\begin{split}\label{events}
&d\dot{\nu}=\frac{1}{2}e^4\;\dbar^3 k_1\;\dbar^3 k_2\;\dbar^3p_+\;\dbar^3p_-\;\deltabar^4_{k_1+k_2,p_++p_-}\\
&\quad \times W_T(\pmb{p}_+) W_T(\pmb{p}_-)\overline{\mathcal{M}^2}_{e^-e^+\to\gamma\gamma}(k_1,k_2,p_+,p_-),
\end{split}
\end{equation}where the $1/2$-factor arises since the emitted  photons of the final state are indistinguishable. For the  sake of simplicity, the  shorthand notations $\deltabar_{p,q}^4=(2\pi)^4\delta^4(p-q)$,  
$\dbar^3k_{1,2} =d^3k_{1,2}/[2\omega_{\pmb{k}_{1,2}}(2\pi)^3]$ and $\dbar^3p_\pm=d^3p_\pm/[2\mathpzc{w}_{\pmb{p}_\pm}(2\pi)^3]$ with $p_\pm=(\mathpzc{w}_{\pmb{p}_\pm},\pmb{p}_\pm)$ and  
$k_{1,2}=(\omega_{\pmb{k}_{1,2}},\pmb{k}_{1,2})$ have been used.  Positive and negative subscripts identify the positron and electron  momentum,  respectively. Notice that 
$\mathpzc{w}_{\pmb{p}_\pm}=[p_{\perp,\pm}^2+p_{\parallel,\pm}^2+m^2]^{\nicefrac{1}{2}}$ is the energy of the positron and the electron when the field has been switched off. 
Here $\overline{\mathcal{M}^2}_{e^-e^+\to\gamma\gamma}$ denotes the unpolarized squared invariant amplitude of the annihilation process [see Fig.~\ref{fig:ext}]. The precise
expression of this object can be found in text books--see for instance Ref.~\cite{Greiner}--and reads 
\begin{equation}
\begin{split}
&\overline{\mathcal{M}^2}_{e^-e^+\to\gamma\gamma}=2\left[\frac{k_2p_-}{k_1p_-}+\frac{k_1p_-}{k_2p_-}\right.\\
&\qquad\qquad\left.+\frac{2m^2 k_1k_2}{(k_1p_-)(k_2 p_-)}-\frac{m^4 (k_1k_2)^2}{(k_1p_-)^2(k_2 p_-)^2}\right].
\end{split}
\end{equation}

At this point it turns out to be convenient to  take  into account the identity $\int \dbar^3k_{i} =\int \frac{d^4k_{i}}{(2\pi)^4}\delta(k_i^2)\Theta(k_{i0})$ and perform the integrations over 
$k_2$ and $\omega_{\pmb{k}_1}$. As a consequence, 
\begin{equation}
\begin{split}
&\frac{d\dot{\nu}}{d\varphi d\theta \sin\theta}=\frac{1}{2}\alpha^2\int \dbar^3p_+\;\dbar^3p_-\;W_T(\pmb{p}_+) W_T(\pmb{p}_-) \\
&\qquad\qquad\qquad \times\frac{(m^2+p_+p_-)\overline{\mathcal{M}^2}_{e^-e^+\to\gamma\gamma}}{(\mathpzc{w}_{\pmb{p}_+}+\mathpzc{w}_{\pmb{p}_-}-\pmb{p}_-\cdot\pmb{n}-\pmb{p}_+\cdot\pmb{n})^2} ,
\end{split}
\end{equation}where $\alpha=1/137$ is the  fine-structure constant. Here  $\theta$ is the  polar angle that the wave vector $\pmb{n}=\pmb{k}_1/{\vert\pmb{k}_1\vert}$ forms with the polarization direction of the  switched off 
field. In contrast,  $\varphi$ represents the azimuthal angle. In the expression above  $\overline{\mathcal{M}^2}_{e^-e^+\to\gamma\gamma}$ must be understood as 
function depending only  on $p_\pm$ and $k_1^\mu=\omega_{\pmb{k}_1}(1,\pmb{n})$ with 
\begin{equation}
\omega_{\pmb{k}_1}=\frac{m^2+p_+p_-}{\mathpzc{w}_{\pmb{p}_+}+\mathpzc{w}_{\pmb{p}_-}-\pmb{p}_-\cdot\pmb{n}-\pmb{p}_+\cdot\pmb{n}}.
\end{equation}

The integrals over $\pmb{p}_{\perp,\pm}$ can also be  carried out approximately. To this end, we first develop the change of variables $\gamma_{\perp,\pm}=(eE_s)^2/[(m^2+\pmb{p}_{\perp,\pm}^2)\omega^2]$ and consider 
the fact that the exponent associated with  $W_T(\pmb{p}_+)$ [$W_T(\pmb{p}_-)$] grows monotonically with $\gamma_{\perp,+}$ [$\gamma_{\perp,-}$]. Hence, after integrating over these variables by parts separately, 
we end up with 
\begin{equation}
\begin{split}\label{integratedevent0}
&\frac{d\dot{\nu}}{d\varphi d\theta \sin\theta}\approx\frac{\alpha^2}{32\pi^4} \left(\frac{E_s}{E_{\mathrm{cr}}}\right)^2 m^4 \int_{-\frac{\pi N}{\gamma}m}^{\frac{\pi N}{\gamma}m}\frac{d p_{\parallel,+}}{2\mathpzc{w}_{p_{\parallel,+}}}\\
&\times\int_{-\frac{\pi N}{\gamma}m}^{\frac{\pi N}{\gamma}m}\frac{d p_{\parallel,-}}{2\mathpzc{w}_{p_{\parallel,-}}}\overline{\mathcal{M}^2}_{e^-e^+\to\gamma\gamma}(p_{\parallel,+},p_{\parallel,-},\theta)\\ 
&\qquad\times\frac{m^2+\mathpzc{w}_{p_{\parallel+}}\mathpzc{w}_{p_{\parallel,-}}+p_{\parallel-}p_{\parallel+}}{[\mathpzc{w}_{p_{\parallel,+}}+\mathpzc{w}_{p_{\parallel,-}}-(p_{\parallel,+}-p_{\parallel,-})\cos\theta]^2}\\
&\qquad\times \frac{\exp\left[-\frac{E_{\mathrm{cr}}}{E_s}\mathrm{Im}\;\mathcal{S}_0(\gamma,p_{\parallel,+})\right]}{[\mathrm{Im}\;\mathcal{S}_0(\gamma,p_{\parallel,+})+\frac{1}{2}\gamma \frac{\partial}{\partial\gamma}\mathrm{Im}\;\mathcal{S}_0(\gamma,p_{\parallel,+})]^2}\\
&\qquad\times \frac{\exp\left[-\frac{E_{\mathrm{cr}}}{E_s}\mathrm{Im}\;\mathcal{S}_0(\gamma,p_{\parallel,-})\right]}{[\mathrm{Im}\;\mathcal{S}_0(\gamma,p_{\parallel,-})+\frac{1}{2}\gamma \frac{\partial}{\partial\gamma}\mathrm{Im}\;\mathcal{S}_0(\gamma,p_{\parallel,-})]^2}.
\end{split}
\end{equation}

The applied procedure manifests  that the main contribution of these integrations results from the region in which $p_{\perp,\pm}/m\ll1$. This result is somewhat expected 
since the single-particle distribution function [see Eqs.~(\ref{distributionfunction}), (\ref{smallkeldych}) and (\ref{asymptoticforW})] is sharply peaked  at $p_\perp=0$, and the typical 
value of  momentum perpendicular to the field direction [$\langle \pmb{p}_\perp\rangle=0$]\footnote{Here $\langle\mathcal{O}\rangle=\int d^3p\; \mathcal{O}(\pmb{p})W_{\infty}(\pmb{p})/\int d^3p\; W_{\infty}(\pmb{p})$. 
For computing the integration with $\mathcal{O}(\pmb{p})=p_\perp^2$,  a procedure similar to the one used for obtaining Eq.~(\ref{main}) has been followed. } 
\begin{equation}
\begin{split}\label{psq}
p_\perp&\sim\langle p_\perp^2\rangle^{\nicefrac{1}{2}}\\
&\approx\frac{m}{\sqrt{\pi}} \left[\frac{E_s}{E_{\mathrm{cr}}}\right]^{\nicefrac{1}{2}} \frac{1}{[1-\mathpzc{g}_\varepsilon(\gamma)-\frac{1}{2}\gamma \mathpzc{g}_\varepsilon^\prime(\gamma)]^{\nicefrac{1}{2}}}\\
\end{split}
\end{equation}remains much smaller than $m$. Hence,  the  photons are emitted quasi-isotropically on the polar plane while their azimutal distribution follows a nontrivial law  to be determined 
in brief. Still, the main trend of this angular distribution can be anticipated when noting that the typical value of momentum parallel to $\pmb{E}_s$  grows linearly with the pulse length [$\langle \pmb{p}_\parallel\rangle=0$]: 
\begin{equation}\label{msvpa}
p_\parallel\sim\langle p_\parallel^2\rangle^{\nicefrac{1}{2}}\approx\frac{1}{2\sqrt{3}} eE_sT.
\end{equation} As this  largely exceeds $p_\perp$ from  Eq.~(\ref{psq}), we find that in average, the angle between the momentum of the created particles and the  external field $\theta\sim \langle p_\perp^2\rangle^{\nicefrac{1}{2}}/\langle p_\parallel^2\rangle^{\nicefrac{1}{2}}\ll1$. Moreover,  Eq.~(\ref{msvpa}) provides evidences that the yielded 
electrons and positrons are mostly ultrarelativistic because $eE_sT/m=2\pi N/\gamma\gg1$. Indeed, their average  energy turns out to be 
\begin{equation}
\langle\mathpzc{w}_{\pmb{p}_\pm}\rangle\approx \frac{1}{4}eE_sT
\end{equation} with a standard deviation  $\Delta \mathpzc{w}_{\pmb{p}_\pm}\approx \pm\langle\mathpzc{w}_{\pmb{p}_\pm}\rangle/\sqrt{3}$. When the energy momentum balance linked to the recombination process 
$p_+^\mu+p_-^\mu=k_1^\mu+k_2^\mu$ is considered, it turns out that--in average--the outcoming photons are emitted back-to-back with a mean energy 
 $\langle\omega_{\pmb{k}_1}\rangle\approx\langle\omega_{\pmb{k}_2}\rangle\approx\langle\mathpzc{w}_{\pmb{p}_\pm}\rangle$. With a strong field strength $E_s\approx 6.6\times 10^{-2} E_{\rm cr}$ and 
a pulse length  $T\sim1\;\rm fs$ corresponding to $N\approx2.1\times 10^5$ for $\omega=1.7\;m$ [$\gamma\approx26$] the mean photon energy is $\langle\omega_{\pmb{k}_1}\rangle\approx6.3\;\rm GeV$.

Returning back to Eq. (\ref{integratedevent0}), the integrals which remain there  cannot be computed analytically. To approximate them, we first develop the change of variables $\gamma_{\parallel,\pm}p_{\parallel,\pm}\omega/(eE_s)$ 
and decompose them in sums over integrals defined over regions with $2\pi-$extensions:
\begin{equation}
\begin{split}
\int_{-\frac{\pi N}{\gamma}m}^{\frac{\pi N}{\gamma}m} d p_{\parallel,\pm}\ldots&=\frac{eE_s}{\omega}\int_{-\pi N}^{\pi N} d \gamma_{\parallel,\pm}\ldots\\
&=\frac{eE_s}{\omega}\sum_{\mathpzc{m}}\int_{(2\mathpzc{m}-1)\pi}^{(2\mathpzc{m}+1)\pi} d \gamma_{\parallel,\pm}\ldots,
\end{split}
\end{equation}where $\mathpzc{m}$ runs  from $\mathpzc{m}_{\mathrm{min}}=-\lfloor(N-1)/2\rfloor$ to $\mathpzc{m}_{\mathrm{max}}=\lfloor(N-1)/2\rfloor$ with $\lfloor x\rfloor$ referring to the integer value of
$x$. Observe that the main contribution of each individual integration results from  the region of $\gamma_{\parallel,i}\in[(2\mathpzc{m}-1)\pi,(2\mathpzc{m}+1)\pi]$ for which the exponent is minimized. Following 
our discussion in Sec.~\ref{SPDFasymptotic}, this takes place at  $\gamma_{\parallel,i}=2\mathpzc{m}\pi$. Hence, we expand each exponent up to  $\sim(\gamma_{\parallel,i}-2\mathpzc{m}\pi)^2$ and set 
$\gamma_{\parallel,i}=2\mathpzc{m}\pi$ in each pre-exponent. Afterwards, $\gamma_{\parallel,\pm}$ are integrated out and we obtain
\begin{equation}
\begin{split}\label{integratedevent}
&\frac{d\dot{\nu}}{d\varphi d\theta \sin\theta}\approx\frac{\alpha^2\pi^2}{(m\omega)^2}\dot{\mathpzc{N}}^2 \mathfrak{F}(\gamma,\theta),
\end{split}
\end{equation} where Eq.~(\ref{main}) has been inserted. Here the function $\mathfrak{F}(\gamma,\theta)$ encodes the angular distribution  in the polar plane and reads
\begin{equation}
\begin{split}
&\mathfrak{F}(\gamma,\theta)=\sum_{\mathpzc{m},\mathpzc{m}^\prime} \frac{Q_{\mathpzc{m},\mathpzc{m}^\prime}}{\mathpzc{E}_{\mathpzc{m}}\mathpzc{E}_{\mathpzc{m}^\prime}[f_{\mathpzc{m}}(\theta)+f_{\mathpzc{m}^\prime}(\theta)]^2}\left[\frac{f_\mathpzc{m}(\theta)}{f_{\mathpzc{m}^\prime}(\theta)}\right.\\
&\qquad\qquad+\frac{f_{\mathpzc{m}^\prime}(\theta)}{f_\mathpzc{m}(\theta)}+\frac{2[f_{\mathpzc{m}}(\theta)+f_{\mathpzc{m}^\prime}(\theta)]^2}{f_\mathpzc{m}(\theta)f_{\mathpzc{m}^\prime}(\theta)Q_{\mathpzc{m},\mathpzc{m}^\prime}}\\
&\qquad\qquad-\left.\frac{[f_{\mathpzc{m}}(\theta)+f_{\mathpzc{m}^\prime}(\theta)]^4}{f_{\mathpzc{m}}(\theta)^2f_{\mathpzc{m}^\prime}(\theta)^2Q_{\mathpzc{m},\mathpzc{m}^\prime}^2}\right].
\end{split}
\end{equation} Other functions  contained in the  expression above, are given by
\begin{equation}
\begin{split}
&Q_{\mathpzc{m},\mathpzc{m}^\prime}=1+\frac{2\mathpzc{m}\pi}{\gamma}\frac{2\mathpzc{m}^\prime\pi}{\gamma}+\mathpzc{E}_{\mathpzc{m}}\mathpzc{E}_{\mathpzc{m}^\prime},\\
&f_{\mathpzc{m}}(\theta)=\mathpzc{E}_{\mathpzc{m}}-\frac{2\mathpzc{m}\pi}{\gamma}\cos\theta,\\
&f_{\mathpzc{m}^\prime}(\theta)=\mathpzc{E}_{\mathpzc{m}^\prime}+\frac{2\mathpzc{m}^\prime\pi}{\gamma}\cos\theta,\\
&\mathpzc{E}_{\mathpzc{m}}=\sqrt{1+\frac{4\pi^2\mathpzc{m}^2}{\gamma^2}},\quad \mathpzc{E}_{\mathpzc{m}^\prime}=\sqrt{1+\frac{4\pi^2\mathpzc{m}^{\prime2}}{\gamma^2}}. 
\end{split}
\end{equation} 

An  estimate for the  number of emitted photons  $\Delta\mathpzc{n}_{\gamma}(t)=\Delta \nu \delta t \ell^3$  in a volume $\ell^3$ can be 
established from Eq.~(\ref{integratedevent}) and reads
\begin{equation}
\begin{split}\label{integratedeventap}
&\Delta\mathpzc{n}_{\gamma}(t)\approx\frac{4\alpha^2\pi^3}{(m\omega)^2}\dot{\mathpzc{N}}\;\mathpzc{n}_{e^-e^+} \frac{\delta t}{T}  \int_{0}^1dx\; \mathfrak{F}(\gamma,x),
\end{split}
\end{equation}where the change of variable $x=\cos\theta$ has been carried out and
\begin{equation}
\begin{split}
&\int_{0}^1dx\; \mathfrak{F}(\gamma,x)=1-\sum_{\mathpzc{m},\mathpzc{m}^\prime\neq0} \frac{Q_{\mathpzc{m},\mathpzc{m}^\prime}}{\mathpzc{E}_{\mathpzc{m}}\mathpzc{E}_{\mathpzc{m}^\prime}}
\left[\frac{2 \mathpzc{G}_{\mathpzc{m},\mathpzc{m}^\prime}}{\mathpzc{E}_{\mathpzc{m}}+\mathpzc{E}_{\mathpzc{m}^\prime}}\right.\\
&\qquad-\left(1+\frac{2}{Q_{\mathpzc{m},\mathpzc{m}^\prime}}+\frac{2}{Q_{\mathpzc{m},\mathpzc{m}^\prime}^2}\right)\frac{\ln\left[\frac{\mathpzc{E}_{\mathpzc{m}}}{\mathpzc{E}_{\mathpzc{m}^\prime}}\frac{f_{\mathpzc{m}^\prime}(1)}{f_{\mathpzc{m}}(1)}\right]}{\frac{2\pi \mathpzc{m}^\prime}{\gamma}\mathpzc{E}_{\mathpzc{m}}+\frac{2\pi\mathpzc{m}}{\gamma}\mathpzc{E}_{\mathpzc{m}^\prime}}\\
&\qquad+\left.\frac{1}{ Q_{\mathpzc{m},\mathpzc{m}^\prime}^2}\left(\frac{1}{\mathpzc{E}_{\mathpzc{m}^\prime} f_{\mathpzc{m}^\prime}(1)}+\frac{1}{\mathpzc{E}_{\mathpzc{m}} f_{\mathpzc{m}}(1)}\right)\right].
\end{split}
\end{equation}Here the function $\mathpzc{G}_{\mathpzc{m},\mathpzc{m}^\prime}$ reads
\begin{equation}
\mathpzc{G}_{\mathpzc{m},\mathpzc{m}^\prime}=\left\{\begin{array}{ccc}
                                                      1&\mathrm{for}&\mathpzc{m}=\mathpzc{m}^\prime,\\ \\ \displaystyle
                                                      \frac{1}{f_{\mathpzc{m}}(1)+f_{\mathpzc{m}^\prime}(1)} &\mathrm{for}&\mathpzc{m}\neq\mathpzc{m}^\prime.
                                                     \end{array}
\right.
\end{equation}It is worth remarking that, to be consistent with our early-time requirement [see below Eq.~(\ref{earlytimes})], 
the  relation  $\Delta\mathpzc{n}_{\gamma}(t)\ll\mathpzc{n}_{e^-e^+}$  has to be satisfied. Observe that such a condition translates into $\delta t\ll\tau$, where the  characteristic time scale ruling 
the perturbative treatment is given by:
\begin{equation}
 \tau=T\left[\frac{4\alpha^2\pi^3}{(m\omega)^2}\dot{\mathpzc{N}}\int_0^1dx\;\mathfrak{F}(\gamma,x)\right]^{-1}\label{tau}.
\end{equation}This  expression can be interpreted as a relaxation time of an electron (positron) till annihilation occurs. We point 
out  that analogous expressions are obtained when elastic scattering processes are considered instead. However, as the differential cross sections linked to such phenomena are comparable with the one associated 
with the  annihilation process, the corresponding time scales should not differ significantly from the one given above. Observe that the $\dot{\mathpzc{N}}^{-1}$ dependence exhibited in Eq.~(\ref{tau}) allows us to 
infer qualitatively the behavior of $\tau$ when the external parameters are variated. For instance, by taking into account the results shown in the left panel of Fig.~\ref{fig:4}, we note that this time scale 
becomes shorter as the field strength $E_s$ increases. Moreover, for a field configuration characterized by $E_s\approx 6.6\times 10^{-2} E_{\rm cr}$, $E_w=10^{-3}E_{\rm cr}$, a pulse length  $T\sim1\;\rm fs$--corresponding to 
$N\approx2.1\times 10^5$ for $\omega=1.7m$--the combined Keldysh parameter is $\gamma\approx26$, $\int_0^1dx\;\mathfrak{F}(\gamma,x)\approx5.5\times 10^4$ and $\dot{\mathpzc{N}}\approx7.4\times 10^{16}\;\mathrm{fs}^{-1}\mu\mathrm{m}^{-3}$. 
Under such a circumstance, the relaxation time amounts to  $\tau\approx1.3\;\rm ns$. This number  exceeds--to put it into relation--the para-positronium lifetime [$\tau_0\approx 0.12\;\rm ns$] by an order of magnitude. 

\begin{figure}
\includegraphics[width=0.55\textwidth]{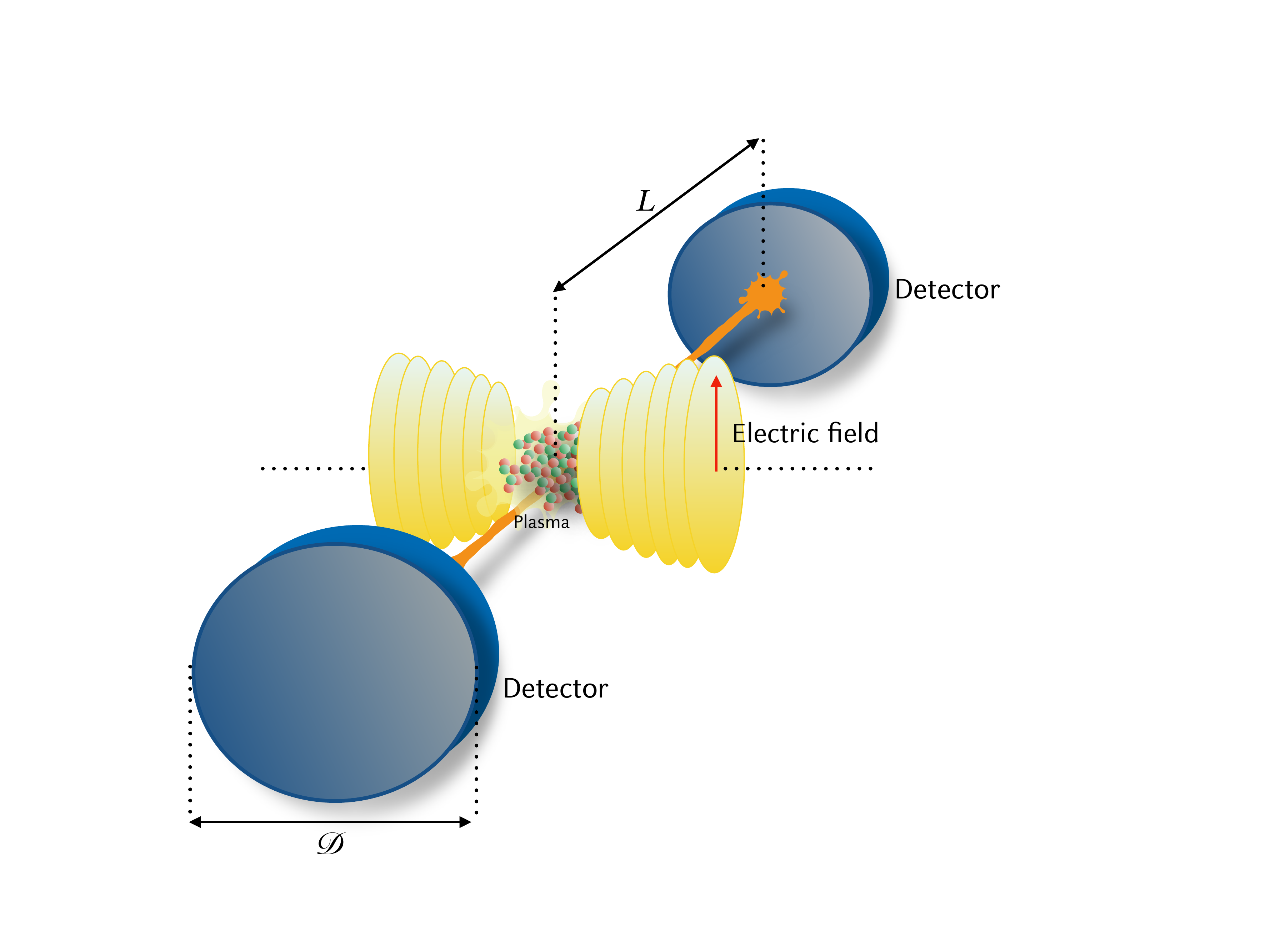}
\caption{Schematic diagram for an experimental setup which probes the possible realization of the dynamical assisted Schwinger mechanism via the emission of high energy photons (in orange) coming out from electron-positron 
annihilations. }
\label{fig:ext2}
\end{figure}

Now, let us  put forward a plausible experimental setup for verifying the dynamically-assisted Schwinger mechanism  via the detection of photons resulting from the  recombination of yielded electron-positron  pairs.  
We propose a scenario in which the high-density electron-positron plasma is generated between two photon detectors, both  placed perpendicular to  the polarization 
direction of the strong field [$\theta=\pi/2$]. We  suppose  both of them  equidistant from the plasma region at a distance $L\approx2\times10^2\;\rm cm$ with their centers forming a right angle  with
respect to the propagation direction of the  colliding pulses [see Fig.~\ref{fig:ext2}]. In addition, we will assume  both detectors characteristized by a length scale $\mathpzc{D}\approx1\;\rm cm$ so that the 
angular openings on the polar and azimuthal planes are very small [$\delta\varphi,\delta\vartheta\approx10^{-2}\;\rm rad$]. With all these details in mind, we can proceed--starting from Eq.~(\ref{integratedevent}) and 
taking into account Eq.~(\ref{tau})--to estimate the amount of photons reaching  a detector due to the recombination after the field has been switched off:
\begin{equation}
\mathpzc{n}_{\mathrm{det}}\sim \mathpzc{n}_{e^-e^+}\frac{\beta\delta\varphi \delta\theta \mathfrak{F}(\gamma,0)}{4\pi\int_0^1dx\;\mathfrak{F}(\gamma,x)},
\end{equation}where the parameter  $\beta=\delta t/\tau$ accounts for the smallness of $\delta t$ relative to $\tau$  [see Eq.~(\ref{tau})]. In order to satisfy safely the early time restriction, we  take $\beta=10^{-2}$, 
i.e. the  measurement time should be $\delta t\approx 10\;\rm  ps$. It is worth remarking that for our reference parameters--see  paragraph that follows Eq.~(\ref{tau})--the form factor attains the value $\mathfrak{F}(\gamma,x=0)\approx 6.9\times 10^3$.  With this details in mind 
we find that $\mathpzc{n}_{\mathrm{det}}\approx 7.4\times 10^{8}$ photons should be detected. Notably, if the weak field is reduced by two orders of magnitude while the remaining parameters keep their values, the initial number of pairs changes to 
$\mathpzc{n}_{e^-e^+}\approx1.2\times10^{12}$ [see red curve in Fig.~\ref{fig:4}] and the number of photons to be detected $\mathpzc{n}_{\mathrm{det}}\approx 1.2\times 10^{4}$ appears still viable experimentally.
 
\section{Conclusions and outlook} \label{sec:numerical}

An analytical investigation of the assisted Schwinger mechanism has been carried out starting from the low density approximation to 
the single-particle distribution function. We have revealed fundamental aspects associated with this process when a weak oscillating field mode 
is superimposed onto a strong constant field background. It has been shown that the particle spectrum is characterized by tiny oscillations 
along the external field direction, whereas perpendicular to it, the spectrum  falls  with the growing of $p_\perp$ significantly more  
slowly than in the case where the production process is driven by the strong field only. Once the field has been switched off, the 
mean-squared values of the momentum depend nontrivialy on the external field parameters, and the quantity linked to the  parallel  component 
to the field exceeds largely the one associated with the perpendicular momentum. As  
a consequence,  most of the particles yielded at that time move approximately parallel to the field direction. Likewise, we have found that the created 
plasma is composed mostly of ultrarelativistic particles and antiparticles.

Both the single-particle distribution function and the density rate of yielded pairs depend on the strong and weak field strenghts in a 
nonperturbative way. While the pair production is predicted to increase significantly in a dynamically-assisted setup, the yielded electron-positron 
plasma has a pronounced unstable nature. This feature demanding to carry out  experimental measurement in  time intervals significantly  
smaller than the plasma life time, otherwise, the beneficial aspect conceded by the assisted setup is lost. In connection, we have 
argued that--under the  early-time  circumstance--the number of photons emitted as a result of electron-positron recombinations 
could be large enough as to constitute an indirect signal of the spontaneous production of pairs from the vacuum. Based on  in this effect, 
a plausible experimental setup for their observation has been put forward. The robustness of our estimates for the number of photons to be 
detected  supports the viability of the proposed  setup as a genuine channel for verifying the Schwinger mechanism,  provided strong field 
strenghts comparable to those to be reached at ELI and XCELS lasers are exploited.

\section*{Acknowledgments}
This work has been funded by the Deutsche Forschungsgemeinschaft (DFG, German Research Foundation) under Grant No. 388720772 (MU 3149/5-1).

\appendix

\section{Sum of integrals over $\pmb{\mathpzc{c}_{\mathpzc{k}}}$ \label{AppendixA}}

This appendix is devoted to determine the contribution due to the poles circumvented by the chosen integration path [see Fig.~\ref{fig:0}]. Explicitly, 
\begin{equation}
\sum_{\mathpzc{k}}\oint_{\mathpzc{c}_{\mathpzc{k}}}\frac{d\tau}{1+\tau^2}\exp\left[\frac{2i\epsilon_\perp^2}{eE_s}\int_0^\tau d\tilde{\tau}\mathpzc{f}(\tilde{\tau})\right],
\end{equation} where the function involved in the exponent is 
\begin{equation}\label{integrandexp}
\mathpzc{f}(\tilde{\tau})=\frac{(1+\tilde{\tau}^2)^{\nicefrac{1}{2}}}{1+\varepsilon \cos(\gamma_\perp \tilde{\tau}-\gamma_\parallel)}.
\end{equation} Over each circle $\mathpzc{c_k}$, we have $\tau=\tau_{+\mathpzc{k}}+\mathpzc{r}/\gamma_\perp e^{i\varphi}$ with $3\pi/2\leqslant\varphi<-\pi/2$ and $\mathpzc{r}$ is 
an infinitesimal quantity [$\mathpzc{r}\to0$]. Consequently,
\begin{equation}
\begin{split}
&\sum_{\mathpzc{k}}\oint_{\mathpzc{c}_{\mathpzc{k}}}\ldots\approx\frac{i\mathpzc{r}}{\gamma_\perp}\sum_{\mathpzc{k}} \frac{1}{1+\tau_{+\mathpzc{k}}^2}\int_{3\pi/2}^{-\pi/2}d\varphi \\
&\qquad\times\exp\left[i\varphi+\frac{2i\epsilon_\perp^2}{eE_s}\int_0^{\tau_{+\mathpzc{k}}+\frac{\mathpzc{r}}{\gamma_\perp}e^{i\varphi}} d\tilde{\tau}\;\mathpzc{f}(\tilde{\tau})\right].
\end{split}\label{polescontribution}
\end{equation} In the following we will show that the  dependence on $\mathpzc{r}$ of the integral over $\varphi$ guarantees that, at  $\mathpzc{r}\to0$, the right-hand side of  Eq.~(\ref{polescontribution}) vanishes. 
To this end, we will focus on determining the leading order dependence  of the integral involved in the exponent. While our exposition will center on the sector in wich $\mathrm{Re}\;\tau_{+\mathpzc{k}}>0$, 
an extension to the remaining case, i.e., $\mathrm{Re}\;\tau_{+\mathpzc{k}}<0$ is straightforward. 

\begin{figure}
\includegraphics[width=0.4\textwidth]{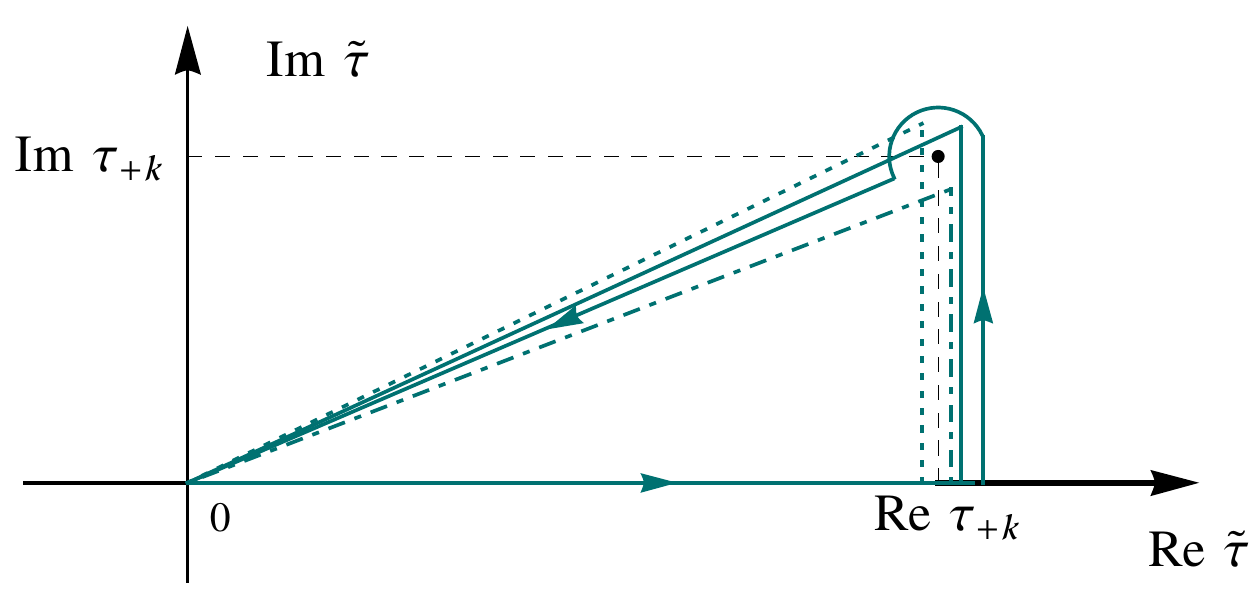}
\caption{Four plausible integration contours chosen to estimate the behavior of the integral involved in the exponent of Eq.~(\ref{polescontribution}) as a function of $\mathpzc{r}$. 
Here we have assumed  $\mathrm{Re}\;\tau_\mathpzc{+k}>0$. The dotted path  applies when the angle $\varphi$ lies in the sector $(\pi/2,3\pi/2]$, whereas the trajectory in  dotdashed 
style is suitable when  $-\pi/2<\varphi<\arg \tau_{+\mathpzc{k}}$. None of these contours  enclose the pole at $\tilde{\tau}=\tau_\mathpzc{+k}$. However, the solid curve in which the 
arc is not present shows a plausible integration path applicable when $\arg \tau_{+\mathpzc{k}} <\varphi<\pi/2$. Conversely, the trajectory including the arc will apply if $\varphi=\arg\tau_{+\mathpzc{k}}$. 
The corresponding integration circuits linked to the case  $\mathrm{Re}\;\tau_{+\mathpzc{k}}<0$ result from reflexions with respect to the imaginary axis [$\mathrm{Re}\;\tilde{\tau}\to-\mathrm{Re}\;\tilde{\tau}$] and 
by taking a counterclockwise sense.}
\label{fig:6}
\end{figure}

The first step toward our aim is to deform appropriately the integration contour of the integral involved in the exponent [see Eq.~(\ref{polescontribution})]. However, 
this procedure  depends on the value of  $\varphi$. Indeed, if $\pi/2<\varphi\leqslant3\pi/2$ or  $-\pi/2<\varphi<\arg \tau_{+\mathpzc{k}}$, the  path can be chosen  
without enclosing the pole $\tau_{+\mathpzc{k}}$. Two plausible circuits, covering the described situation are shown Fig.~\ref{fig:6} in dotted and dotdashed styles. 
Conversely, if $\arg \tau_{+\mathpzc{k}} \leqslant\varphi\leqslant\pi/2$, the contour will be chosen containing the pole inside the trajectory. Observe that those cases 
in which $\varphi=\arg \tau_{+\mathpzc{k}}$ or $\varphi =\pi/2$ demand to surround the pole via a small arc of radius $\mathpzc{r}/\gamma_\perp$ [see Fig.~\ref{fig:6}]. 
Obviously, the situation described previously implies that the integral over $\varphi$ has to be splitted into three portions; each of which covering 
one of the described  $\varphi$-sectors:
\begin{equation}
\begin{split}\label{splittingouterintegra1}
&\int_{3\pi/2}^{-\pi/2} d\varphi\ldots=\lim_{\eta_1,\eta_2\to0^+}\left[\int_{3\pi/2}^{\pi/2+\eta_1} d\varphi\ldots\right.\\
&\qquad+\left.\int^{\arg\;\tau_{+\mathpzc{k}}+\eta_2}_{\pi/2-\eta_1} d\varphi\ldots+\int_{\arg\;\tau_{+\mathpzc{k}}-\eta_2}^{-\pi/2} d\varphi\ldots\right].
\end{split}
\end{equation}

Let us consider the integration over the path enclosing the pole via the arc [see Fig.~\ref{fig:6}]. As a consequence of the residue theorem [$\mathrm{Re}\;\tau_{+\mathpzc{k}}>0$]: 
\begin{equation}
\begin{split}\label{residueapplication}
&\int_0^{\tau_{+\mathpzc{k}}+\frac{\mathpzc{r}}{\gamma_\perp}e^{i(\pi+\arg\tau_{+\mathpzc{k}})}} d\tilde{\tau}\;\mathpzc{f}(\tilde{\tau})=2\pi i\mathrm{Res}[\tau_{+\mathpzc{k}},\mathpzc{f}(\tau)]\\
&\qquad\qquad\qquad\qquad+\mathpzc{J}_{\mathrm{arc}}(\mathpzc{r})+\mathpzc{J}_{\mathrm{Re}\;\tau}(\mathpzc{r})+\mathpzc{J}_{\mathrm{ver}}(\mathpzc{r}),\\
&\mathpzc{J}_{\mathrm{arc}}(\mathpzc{r})=\frac{i\mathpzc{r}}{\gamma_\perp}\int_{\arg\tau_{+\mathpzc{k}}}^{\pi+\arg\tau_{+\mathpzc{k}}}d\phi\; e^{i\phi}\mathpzc{f}\left(\tau_{+\mathpzc{k}}+\frac{\mathpzc{r}}{\gamma_\perp}e^{i\phi}\right),\\
&\mathpzc{J}_{\mathrm{Re}\;\tau}(\mathpzc{r})=\int_0^{\mathrm{Re}\;\tau_{+\mathpzc{k}}+\frac{\mathpzc{r}}{\gamma_\perp}\cos(\arg\tau_{+\mathpzc{k}})}dx\;\mathpzc{f}(x),\\
&\mathpzc{J}_{\mathrm{ver}}(\mathpzc{r})=i\int_0^{\mathrm{Im}\;\tau_{+\mathpzc{k}}+\frac{\mathpzc{r}}{\gamma_\perp}\sin(\arg\tau_{+\mathpzc{k}})}dy\;\\
&\qquad\qquad\times\mathpzc{f}\left(\mathrm{Re}\;\tau_{+\mathpzc{k}}+\frac{\mathpzc{r}}{\gamma_\perp}\cos(\arg\tau_{+\mathpzc{k}})+iy\right).
\end{split}
\end{equation}In those cases in which the integration contour avoids the pole, the Cauchy theorem applies and the integral of interest, i.e. the left-hand side in the first line of Eq.~(\ref{residueapplication}) 
with $\pi+\arg\tau_{+\mathpzc{k}}\to\varphi$, is determined by two contributions similar to $\mathpzc{J}_{\mathrm{Re}\;\tau}(\mathpzc{r})$ and $\mathpzc{J}_{\mathrm{ver}}(\mathpzc{r})$, with 
$\arg\tau_{+\mathpzc{k}}\to \varphi$ and $\varphi$ taking values within the respective sector [see below Eq.~(\ref{polescontribution})]. Conversely, if the  inclusion of the  pole is required  without 
the necessity of a circumventing arc, the expression for the integral involved in the exponent will coincide with Eq.~(\ref{residueapplication}) upto $\mathpzc{J}_{\mathrm{arc}}(\mathpzc{r})$, provided 
the replacement  $\pi+\arg\tau_{+\mathpzc{k}}\to\varphi$  in the first line of Eq.~(\ref{residueapplication}) is carried out. Likewise, one will be forced to replace $\arg\tau_{+\mathpzc{k}}\to \varphi$ 
in both $\mathpzc{J}_{\mathrm{Re}\;\tau}(\mathpzc{r})$ and $\mathpzc{J}_{\mathrm{ver}}(\mathpzc{r})$. Thus, the analysis of the expression  above allows us to infer the outcomes related to the diverse 
$\varphi$ values.

Both the residue  of the function  $\mathpzc{f}(\tilde{\tau})$ at $\tau_{+\mathpzc{k}}$  as well as the leading order contribution linked to the tiny arc in Fig.~\ref{fig:6}:
\begin{equation}
\mathpzc{J}_{\mathrm{arc}}(\mathpzc{r}) \stackrel{\mathpzc{r}\to0}{\approx}-\frac{\pi}{\gamma_\perp}(1+\tau_{+\mathpzc{k}}^2)^{\nicefrac{1}{2}}
\end{equation}are independent of $\mathpzc{r}$. 
Regarding the  behavior of the integral $\mathpzc{J}_{\mathrm{Re}\;\tau}(\mathpzc{r})$, i.e., the third contribution in the right-hand side of Eq.~(\ref{residueapplication}); 
here the oscillatory contribution present in its integrand is always much smaller than unity. As a consequence, one can ignore its effect by approaching $[1+\varepsilon \cos(\gamma_\perp x-\gamma_\parallel)]^{-1}\approx1$  
and the integral 
\begin{equation}
\begin{split}
\lim_{\mathpzc{r}\to0}&\mathpzc{J}_{\mathrm{Re}\;\tau}(\mathpzc{r})\approx\frac{1}{2}\Big[ \mathrm{Re}\;\tau_{+\mathpzc{k}}\sqrt{1+[\mathrm{Re}\;\tau_{+\mathpzc{k}}]^2}\\
&\qquad\qquad+\sin^{-1}(\mathrm{Re}\;\tau_{+\mathpzc{k}})\Big]
\end{split}\label{sigto}
\end{equation}with $\sqrt{1+[\mathrm{Re}\;\tau_{+\mathpzc{k}}]^2}\geqslant0$ becomes independent of  $\mathpzc{r}$ as  $\mathpzc{r}\to0$. 

Now we focus on the last integral  $\mathpzc{J}_{\mathrm{ver}}(\mathpzc{r})$ in Eq.~(\ref{residueapplication}), the calculation of which requires a somewhat elaborate procedure. Mainly, because  its dependence 
on  $\mathrm{Im}\;\tau_{+\mathpzc{k}}$ leads  to analyze the regimes $\mathrm{Im}\;\tau_{+\mathpzc{k}}\ll1$ and  $\mathrm{Im}\;\tau_{+\mathpzc{k}}\gg1$ separately. As this is the only plausible contribution 
which may  provide a nontrivial  dependence on $\mathpzc{r}$, we will replace in it $\arg\tau_{+\mathpzc{k}}$ by $\varphi$,  and consider $\varphi$ taking values within the integration region in Eq.~(\ref{polescontribution}).

\subsection{The case of strong enhancement: $\pmb{\mathrm{Im}\;\tau_{+\mathpzc{k}}\ll1}$}

Let us consider first the situation in which  $\mathrm{Im}\;\tau_{+\mathpzc{k}}\ll1$.\footnote{Note that $\mathrm{Im}\;\tau_{+\mathpzc{k}}=\gamma_{\rm cr}/\gamma_\perp$ is actually independent of $\mathpzc{k}$.} 
Because of this, the square root involved in the integrand of $\mathpzc{J}_{\mathrm{ver}}(\mathpzc{r})$ is approximately  independent  of $y\equiv\mathrm{Im}\;\tilde{\tau}$: $(1+[\mathrm{Re}\;\tau_{+\mathpzc{k}}+iy]^2)^{\nicefrac{1}{2}}\approx\sqrt{1+[\mathrm{Re}\;\tau_{+\mathpzc{k}}]^2}$. 
Indeed, when the  condition $\mathrm{Im}\;\tau_{+\mathpzc{k}}\ll1$  holds,  one can exploit the fact that $y$ is much smaller than unity [$1\gg y$]. 
Observe that this also  implies  that $\arg(1+[\mathrm{Re}\;\tau_{+\mathpzc{k}}+iy]^2)\approx 0$, for all allowed values of $\mathrm{Re}\;\tau_{+\mathpzc{k}}$. Consequently,
\begin{equation}
\begin{split}\label{jintegration}
&\mathpzc{J}_{\mathrm{ver}}(\mathpzc{r})\approx i\sqrt{1+[\mathrm{Re}\;\tau_{+\mathpzc{k}}]^2}\int_0^{\mathrm{Im}\;\tau_{+\mathpzc{k}}+\frac{\mathpzc{r}}{\gamma_\perp}\sin\varphi}dy\\
&\qquad\qquad\qquad\quad\times\frac{1}{1-\varepsilon\cosh(\gamma_\perp y-i\mathpzc{r}\cos\varphi)}.
\end{split}
\end{equation}At this point it turns out to be very convenient to perform the change of variable $s=\exp[\gamma_\perp y]$.  
The calculation of the resulting integral  is simplified once its integrand is decomposed into partial fractions.  The described procedure leads to  write
\begin{equation}
\begin{split}
&\int_0^{\mathrm{Im}\;\tau_{+\mathpzc{k}}+\frac{\mathpzc{r}}{\gamma_\perp}\sin\varphi} dy\ldots\stackrel{\mathpzc{r}\to0}{\sim}-\frac{1}{\gamma_\perp}(\ln\mathpzc{r}+i\varphi),
\end{split}\label{seconintgerationpole}
\end{equation}where an unessential imaginary term independent of $\mathpzc{r}$ and $\varphi$ has been ommited. Inserting Eq.~(\ref{seconintgerationpole}) into Eq.~(\ref{jintegration}), 
we find that the essential contribution of the integral involved in the exponent is
\begin{equation}\label{fjintegration}
\begin{split}
&\int_0^{\tau_{+\mathpzc{k}}+\frac{\mathpzc{r}}{\gamma_\perp}e^{i\varphi}} d\tilde{\tau}\;\mathpzc{f}(\tilde{\tau})\stackrel{\mathpzc{r}\to0}{\sim}- \frac{i}{\gamma_\perp}\\
&\qquad\qquad\qquad\times\sqrt{1+[\mathrm{Re}\;\tau_{+\mathpzc{k}}]^2}(\ln\mathpzc{r}+i\varphi).
\end{split}
\end{equation}Correspondingly, the outer integration in Eq.~(\ref{polescontribution}) [see also Eq.~(\ref{splittingouterintegra1})] behaves as
\[
\int_{3\pi/2}^{-\pi/2}d\varphi\ldots\stackrel{\mathpzc{r}\to0}{\sim}\mathpzc{r}^{\frac{2\epsilon_\perp}{\omega}\sqrt{1+[\mathrm{Re}\;\tau_{+\mathpzc{k}}]^2}},
\] which guarantees that--in the regime particularized by the strong enhancement [$\mathrm{Im}\;\tau_{+\mathpzc{k}}\ll1$]--the sum over the circles eluding the poles gives no contribution to the single-particle 
distribution function $W_T(\pmb{p})$,  provided the appropriate limit  $\mathpzc{r}\to0$ is taken.

\subsection{The case of weak enhancement: $\pmb{\mathrm{Im}\;\tau_{+\mathpzc{k}}\gg1}$}

The  pole  always lies  below the line $\mathpzc{C_R}$ and to the left of $\mathpzc{C_+}$ [read also discussion above Eq.~(\ref{sisi})]. This provides the following condition for the real  and imaginary parts 
of the pole $\tau_{+\mathpzc{k}}$:
\begin{equation}
\begin{split}
&\sqrt{3}\left[\mathrm{Im}\;\tau_{+\mathpzc{k}}-1\right]<\mathrm{Re}\;\tau_{+\mathpzc{k}}<\frac{\pi N}{\gamma_\perp},\\
&\mathrm{Im}\;\tau_{+\mathpzc{k}}<\frac{1}{\sqrt{3}}\mathrm{Re}\;\tau_{+\mathpzc{k}}+1<\frac{1}{\sqrt{3}}\frac{\pi N}{\gamma_\perp}.
\end{split}
\end{equation} However, here we will restrict ourselves to the case in which $\mathrm{Im}\;\tau_{+\mathpzc{k}}\gg1$. As a consequence,  $\mathrm{Re}\;\tau_{+\mathpzc{k}}\gg1$ and the square root involved in the integrand of $\mathpzc{J}_{\mathrm{ver}}(\mathpzc{r})$ [see Eq.~(\ref{residueapplication})] 
behaves as $(1+[\mathrm{Re}\;\tau_{+\mathpzc{k}}+iy]^2)^{\nicefrac{1}{2}}\approx \mathrm{Re}\;\tau_{+\mathpzc{k}}+i y$ and
\begin{equation}
\mathpzc{J}_{\mathrm{ver}}(\mathpzc{r})\approx i\int_0^{\mathrm{Im}\;\tau_{+\mathpzc{k}}+\frac{\mathpzc{r}}{\gamma_\perp}\sin\varphi}\frac{dy( \mathrm{Re} \;\tau_{+\mathpzc{k}}+i y)}{1-\varepsilon\cosh(\gamma_\perp y-i\mathpzc{r}\cos\varphi)}.
\end{equation}The part  of the integration that contains $\mathrm{Re} \;\tau_{+\mathpzc{k}}$ can be calculated following the procedure described to determine the integral in Eq.~(\ref{jintegration}). 
Taking into account Eq.~(\ref{seconintgerationpole}),
\begin{equation}
\begin{split}\label{sssz}
&\mathpzc{J}_{\mathrm{ver}}(\mathpzc{r})\stackrel{\mathpzc{r}\to0}{\sim} -\frac{i}{\gamma_\perp}\mathrm{Re} \;\tau_{+\mathpzc{k}}\left(\ln\mathpzc{r}+i\varphi\right)\\
&\qquad\quad -\frac{1}{\gamma_\perp^2}\int_1^{e^{\gamma_{\mathrm{cr}}+\mathpzc{r}\sin\varphi}}ds\left[\frac{\ln(s)}{s-s_+}-\frac{\ln(s)}{s-s_-}\right],
\end{split}
\end{equation}where the change of variable $s=\exp[\gamma_\perp y]$ has been used. Here $s_+=\frac{2}{\varepsilon}\exp[i \mathpzc{r}\cos\varphi]$ and $s_-=\frac{\varepsilon}{2}\exp[i\mathpzc{r}\cos\varphi]$.

The main contribution to the first  integral in Eq.~(\ref{sssz}) results from $s\sim s_+$. Therefore, up to an unessential imaginary term independent of $\mathpzc{r}$ and $\varphi$, we find 
\begin{equation}
\begin{split}\label{sssz1}
&\frac{1}{\gamma_\perp^2}\int_1^{e^{\gamma_{\mathrm{cr}}+\mathpzc{r}\sin\varphi}}ds\frac{\ln(s)}{s-s_+}\stackrel{\mathpzc{r}\to0}{\approx}\mathrm{Im}\;\tau_{+\mathpzc{k}}\\
&\qquad\qquad\qquad\times\left[\mathrm{Im}\;\tau_{+\mathpzc{k}}+\frac{1}{\gamma_\perp}\left(\ln\mathpzc{r}+i\varphi \right)\right].
\end{split}
\end{equation}Conversely, the integration variable linked to the last integral in Eq.~(\ref{sssz}) satisfies the condition $s\geqslant1\gg s_-$. Hence, by safely ignoring $s_-$ in the  denominator,  we end up with
\begin{equation}\label{sssz2}
\begin{split}
&\lim_{\mathpzc{r}\to0}\frac{1}{\gamma_\perp^2}\int_1^{e^{\gamma_{\mathrm{cr}}+\mathpzc{r}\sin\varphi}}ds\frac{\ln(s)}{s-s_-}\approx\frac{1}{2}[\mathrm{Im}\;\tau_{+\mathpzc{k}}]^2.
\end{split}
\end{equation}Combining the outcomes in Eq.~(\ref{sssz1}) and (\ref{sssz2}) into Eq.~(\ref{sssz}), we obtain
\begin{equation}
\begin{split}
\mathpzc{J}_{\mathrm{ver}}(\mathpzc{r})\stackrel{\mathpzc{r}\to0}{\sim}&-\frac{1}{\gamma_\perp}\left[\mathrm{Im}\;\tau_{+\mathpzc{k}}+i\mathrm{Re}\;\tau_{+\mathpzc{k}}\right]\ln\mathpzc{r} \\
&+\frac{1}{\gamma_\perp}\left[\mathrm{Re}\;\tau_{+\mathpzc{k}}-i\mathrm{Im}\;\tau_{+\mathpzc{k}}\right]\varphi.
\end{split}
\end{equation} Inserting this result into Eq.~(\ref{residueapplication}), it is straightforward to verify that the outer  integration in Eq.~(\ref{polescontribution}) behaves as
\begin{equation}
\int_{3\pi/2}^{-\pi/2}d\varphi\ldots \stackrel{\mathpzc{r}\to0}{\sim}\mathpzc{r}^{\frac{2\epsilon_\perp}{\omega}\mathrm{Re}\;\tau_{+\mathpzc{k}}}\left(\cos\vartheta-i\sin\vartheta\right)
\end{equation}with $\vartheta=2\epsilon_\perp \omega^{-1}\ln\vert\mathpzc{r}\vert\mathrm{Im}\tau_{+\mathpzc{k}}$.  Therefore, in the limit of $\mathpzc{r}\to 0$ and under the condition $\mathrm{Im}\;\tau_{+\mathpzc{k}}\gg1$, 
the sum over the circles $\mathpzc{c_k}$ [see Eq.~(\ref{polescontribution})] provides no contribution either to the single-particle distribution function $W_T(\pmb{p})$.

\begin{figure}
\includegraphics[width=0.4\textwidth]{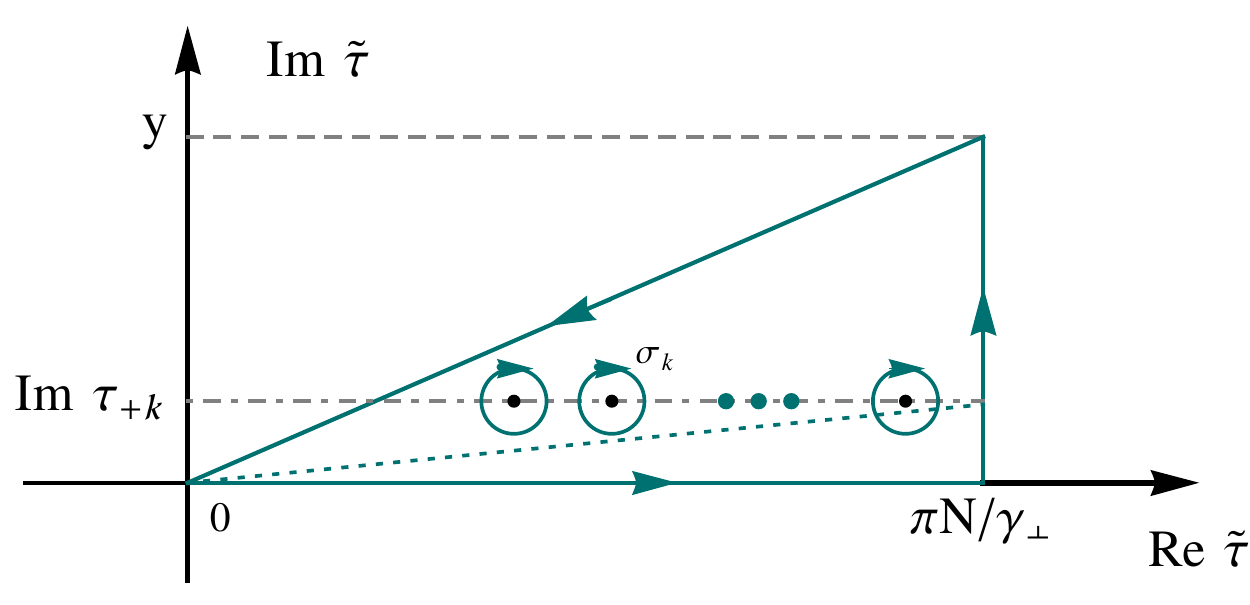}
\caption{Integration contours  chosen to estimate the behavior of $\mathcal{J}_{+}$ [see Eq.~(\ref{sszh10})] in terms of the number of cycles $N$. While the thick trajectory 
is suitable for $y>\mathrm{Im}\;\tau_{+\mathpzc{k}}$, the path in dotted style is appropriate if $y<\mathrm{Im}\;\tau_{+\mathpzc{k}}$. The corresponding integration circuits 
linked to  $\mathcal{J}_{-}$ result from reflexions with respect to the imaginary axis [$\mathrm{Re}\;\tilde{\tau}\to-\mathrm{Re}\;\tilde{\tau}$] and by taking a counterclockwise 
sense. Shortcuts joining each circumventing circle  with the real $\tilde{\tau}-$axis have been omitted for simplicity. As in Fig.~\ref{fig:0}, they give no contribution as both 
lie infinitesimally close together and have opposite orientation.}
\label{fig:7}
\end{figure}

\section{No contribution over $\pmb{\mathpzc{C}_{\pm}}$ \label{AppendixB}}

In this appendix we show that the integrals over the vertical segments $\mathpzc{C}_{\pm}$ [see Fig.~\ref{fig:0}] give no contribution in 
the limit $N\to\infty$. Let us first denote
\begin{equation}
\begin{split}
&I_{\pm}=\int_{\mathpzc{C_{\pm}}}\frac{d\tau}{1+\tau^2}\exp\left[\frac{2i\epsilon_\perp^2}{eE_s}\int_0^\tau d\tilde{\tau}\mathpzc{f}(\tilde{\tau})\right].
\end{split}
\end{equation} The function contained in the exponent can be found in Eq.~(\ref{integrandexp}). Along $\mathpzc{C_+}$, the complex integration variable is characterized 
by  $\tau=\pi N/\gamma_\perp+iy$, whereas over $\mathpzc{C_-}$ by  $\tau=-\pi N/\gamma_\perp+iy$. Over the former trajectory the minimum and maximum values of $y$ 
are $y_{\mathrm{min}}=0$ and $y_{\mathrm{max}}=\pi N/(\sqrt{3}\gamma_\perp)+1$. Conversely, over the latter these values exchange their roles. For very large value of  
$N\gg1$, 
\begin{equation}
\begin{split}\label{sszh10}
&\vert I_{\pm}\vert <\frac{\gamma_\perp^2}{\pi^2 N^2}\int_{0}^{y_{\mathrm{max}}} dy \exp\left[-\frac{\epsilon_\perp^2}{2eE_s}\mathrm{Im}\;\mathcal{J}_{\pm}\right],\\
&\mathcal{J}_{\pm}\equiv\mathcal{J}\left(\pm \frac{\pi N}{\gamma_\perp}+iy\right)=\int_0^{\pm\frac{\pi N}{\gamma_\perp}+iy} d\tilde{\tau}\mathpzc{f}(\tilde{\tau}).
\end{split}
\end{equation}

\begin{figure}
\includegraphics[width=0.45\textwidth]{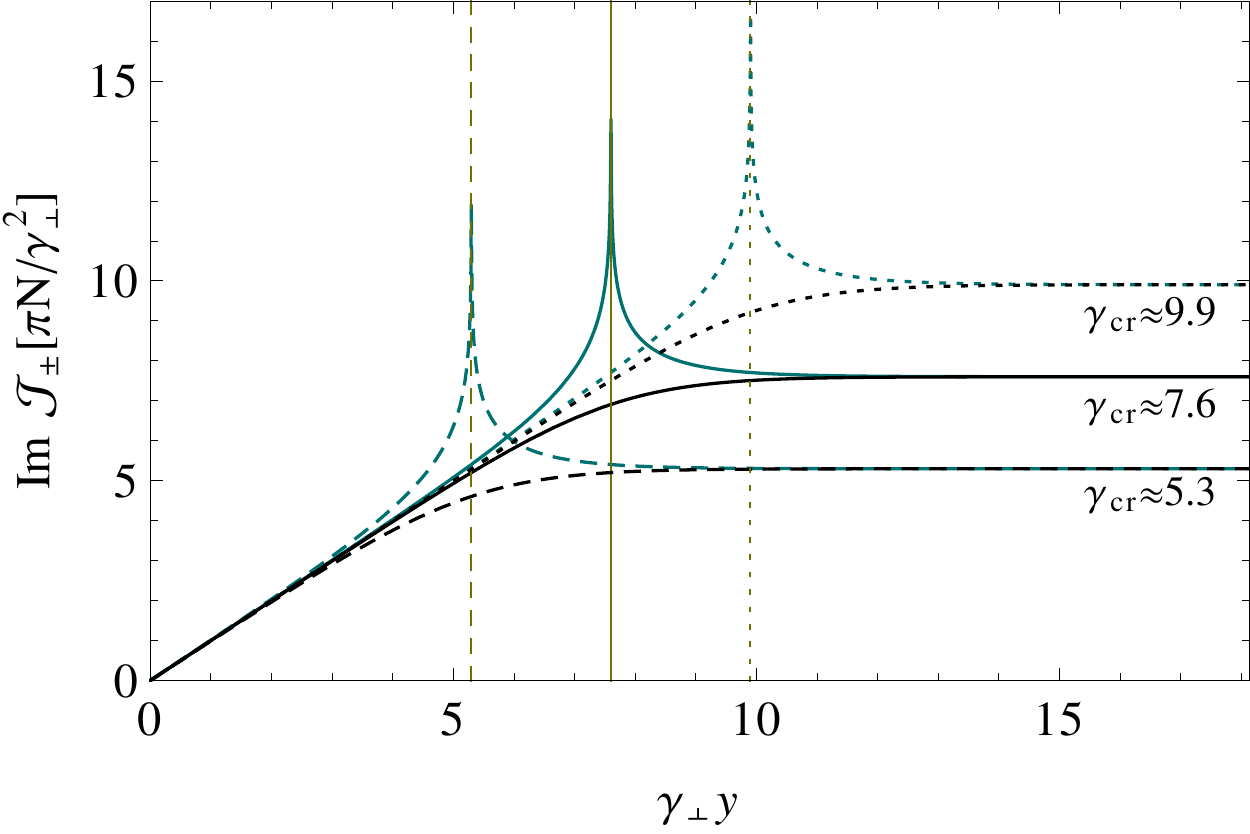}
\caption{Behavior of the integral in Eq.~(\ref{puntoantes}). The curves have been obtained for $\varepsilon=10^{-2}$ (dashed), $\varepsilon=10^{-3}$ (solid) and  $\varepsilon=10^{-4}$ (dotted). 
The vertical dashed line in olive  depicts the value  $\gamma_{\mathrm{cr}}\approx5.3$, the solid one corresponds to $\gamma_{\mathrm{cr}}\approx7.6$, whereas $\gamma_{\mathrm{cr}}\approx9.9$ is shown 
in dotted style.  While the curves in black have been obtained for an even number of cycles, those in darker cyan follow for an odd value of $N$. Each curve manifests a monotonic growth for 
$0\leqslant y\leqslant\gamma_{\mathrm{cr}}/\gamma_\perp$. For $y>\gamma_{\mathrm{cr}}/\gamma_\perp$, the curves in darker cyan--corresponding to an odd $N$--fall very fast toward  the respective  
values of $\gamma_{\mathrm{cr}}$. In contrast,  each curve in black tends smoothly to the due  $\gamma_{\mathrm{cr}}$ parameter.}
\label{fig:8}
\end{figure}

In the following the integrals $\mathcal{J}_{\pm}$ are evaluated by modifying the corresponding  paths.  However, these defomartions depend 
upon whether the outer integration variable $y$ is smaller or greater than $\mathrm{Im}\;\tau_{+\mathpzc{k}}$. Because of this, it turns out to be  beneficial 
to introduce a  positive parameter $y_0\to0^+ $ and split the integral over $y$ accordingly
\begin{equation}
\begin{split}\label{splittingouterintegra}
&\int_{0}^{y_{\mathrm{max}}} dy\ldots=\lim_{y_0\to0^+}\left[\int_{0}^{\mathrm{Im}\;\tau_{+\mathpzc{k}}-y_0} dy\ldots\right.\\
&\qquad\qquad\qquad\qquad\qquad\qquad+\left.\int_{\mathrm{Im}\;\tau_{+\mathpzc{k}}+y_0}^{y_{\mathrm{max}}} dy\ldots\right].
\end{split}
\end{equation}
Within the  sector $\mathrm{Im}\;\tau_{+\mathpzc{k}}+y_0 \leqslant y\leqslant y_{\mathrm{max}}$--covered by the second integral in the right-hand side of Eq.~(\ref{splittingouterintegra})--the chosen 
integration circuit exhibited in  Fig.~\ref{fig:7} avoids the poles at $\tau_{+\mathpzc{k}}$ with $\mathpzc{k}\in \mathbb{Z}$ [see Eq.~(\ref{poles})]. Conversely, if the outer integration  variable 
lies within  $0\leqslant y\leqslant\tau_{+\mathpzc{k}}-y_0$, we can integrate  $\mathcal{J}_{1,4}$--inside the first contribution of the right-hand side in Eq.~(\ref{splittingouterintegra})--via  a 
similar trajectory,  with the particularization that its vertical path ends at $y\leqslant\tau_{+\mathpzc{k}}-y_0$, returning to the origin afterward [dotted line in Fig.~\ref{fig:7}]. As all poles 
are located above the dotted line, no circumvention of them is required in this case.

We will focus on the result coming from the path in which the poles are eluded. However, we will  extract parallely the outcomes linked to the contour in which these 
circumventions are not necessary. The next step in our analysis is the  application of the Cauchy's theorem to the problem under consideration. Correspondingly,   
\begin{equation}
\begin{split}
\mathcal{J}_{\pm}&=\int_0^{\pm\frac{\pi N}{\gamma_\perp}}dx\;\mathpzc{f}(x)+i\int_0^{y} d\tilde{y}\;\mathpzc{f}(\pm\pi N/\gamma_\perp+i\tilde{y})\\
&+\sum_{\mathpzc{k}}\oint_{\sigma_\mathpzc{k}}d\tilde{\tau}\mathpzc{f}(\tilde{\tau}).
\end{split}
\end{equation}
When  $0\leqslant y\leqslant\tau_{+\mathpzc{k}}-y_0$ the expression for $\mathcal{J}_{\pm}$ is only given by the first line of this formula. We note that the integral over the $\mathrm{Re}\;\tilde{\tau}-$axis is purely 
real as well as the leading order contribution of each integration over the small circle $\sigma_{\mathpzc{k}}$ [$\tau=\tau_{+\mathpzc{k}}+\mathpzc{r}/\gamma_\perp e^{i\chi}$ with $3\pi/2\leqslant\chi<-\pi/2$]:  
\begin{equation}
\begin{split}
&\oint_{\sigma_\mathpzc{k}}d\tilde{\tau}\mathpzc{f}(\tilde{\tau})\stackrel{\mathpzc{r}\to0}{\approx}\frac{2\pi}{\gamma_\perp}\left(1+\tau_{+\mathpzc{k}}^2\right)^{\nicefrac{1}{2}}.
\end{split}
\end{equation} Therefore, neither  $\int_0^{\pm\frac{\pi N}{\gamma_\perp}}dx\;\mathpzc{f}(x)$  nor the involved sum contribute to $\mathrm{Im}\;\mathcal{J}_{\pm}$,  which is precisely  what we need  
[see Eq.~(\ref{sszh10})]. In the second integration,  we can exploit the condition $\pi N/\gamma_\perp\gg1$ to approximate the multivalued function 
$\left[1+(\pi N/\gamma_\perp\pm iy)^2\right]^{\nicefrac{1}{2}}\approx \pi N/\gamma_\perp\pm iy$. Consequently,
\begin{equation}
\begin{split}\label{puntoantes}
&\mathrm{Im}\;\mathcal{J}_\pm\approx\mathrm{Im}\;i\int_0^{y} d\tilde{y}\;\mathpzc{f}(\pm\pi N/\gamma_\perp+i\tilde{y})\\
&\qquad\stackrel{N\to\infty}{\approx}\frac{\pi N}{\gamma_\perp}\int_0^{y} d\tilde{y}\frac{1}{1+\varepsilon(-1)^N\cosh(\gamma_\perp\tilde{y})}.
\end{split}
\end{equation}The integral involved in this expression belongs to the class  of integrals  treated previously in appendix~\ref{AppendixA} and can be calculated straightforwardly. This leads to write
\begin{equation}
\begin{split}\label{importanbehaviorintegral}
&\mathrm{Im}\;\mathcal{J}_\pm\approx \frac{\pi N}{\gamma_\perp^2}\left[y\gamma_\perp-\frac{1}{\gamma_\perp}\ln\left\vert 1+\frac{\varepsilon}{2}(-1)^N e^{\gamma_\perp y}\right\vert\right].
\end{split}
\end{equation}

With this expression to our disposal,  the estimation of the integral defined over the sector $0\leqslant y\leqslant \mathrm{Im}\;\tau_{+\mathpzc{k}}-y_0$ [see Eq.~(\ref{splittingouterintegra})] can be carried 
out without much difficulties. At this point, it is worth mentioning that its integrand has an exponent whose absolute value  grows monotonically with  $y$ [see Fig.~\ref{fig:8}]. Hence, the main contribution 
of $\int_0^{\tau_{+\mathpzc{k}}-y_0}dy\ldots$ results from those values  $y\ll \mathrm{Im}\;\tau_{+\mathpzc{k}}-y_0$. Accordingly, $\mathrm{Im}\;\mathcal{J}_\pm\approx\pi N y/\gamma_\perp$ and 
\begin{equation}
\begin{split}\label{szi}
&\lim_{y_0\to 0^+}\int_0^{\mathrm{Im}\;\tau_{+\mathpzc{k}}-y_0}dy\ldots\approx\frac{2eE_s}{\epsilon_\perp^2}\frac{\gamma_\perp}{\pi N}\\
&\qquad\qquad\qquad\qquad\qquad\times\left[1-\left(\frac{\varepsilon}{2}\right)^{\frac{eE_s}{2\omega^2}\pi N}\right].
\end{split}
\end{equation}Contrary to the previous case, the integral defined over  $\mathrm{Im}\;\tau_{+\mathpzc{k}}+y_0 \leqslant y\leqslant y_{\mathrm{max}}$ depends on whether the number of cycles is odd or even. In the former case  
the absolute value of the exponent decreases sharply with the increasing of $y$ toward the value  $\sim \epsilon_\perp^2\mathcal{J}_{\pm}/(2eE_s)$ with $\mathcal{J}_{\pm}\approx\pi N \gamma_{\mathrm{cr}}/\gamma_\perp^2$. 
However, in the latter situation [$N$ even] there exist an almost unappreciable growing toward the same value as the integration variable $y$ increases [see Fig.~\ref{fig:8}]. Hence, the area below both curves approaches
\begin{equation}
\begin{split}\label{szf}
&\lim_{y_0\to 0^+}\int_{\mathrm{Im}\;\tau_{+\mathpzc{k}}+y_0}^{y_{\mathrm{max}}}dy\ldots\approx\left(\frac{\varepsilon}{2}\right)^{\frac{eE_s}{2\omega^2}\pi N}\\
&\qquad\qquad\qquad\qquad\times\left[\frac{\pi N}{\sqrt{3}\gamma_\perp}-\mathrm{Im}\;\tau_{+\mathpzc{k}}\right].
\end{split}
\end{equation} Observe that the insertion of  Eqs.~(\ref{szi}) and (\ref{szf}) into Eq.~(\ref{splittingouterintegra}) ensures the convergence to zero of the right-hand side of the first line in  Eq.~(\ref{sszh10}) 
as $N\to\infty$. Therefore, no contribution over $\mathpzc{C_\pm}$ arises to  the single-particle distribution function $W_T(\pmb{p})$. We remark that this statement applies independently of the  value of 
$\mathrm{Im}\;\tau_{+\mathpzc{k}}$.

\end{document}